\documentclass[12pt,preprint]{aastex}

\shorttitle{Turbulence Cascade}
\shortauthors{Jiang et al.}

\begin{document}

\title{Cascade and Damping of Alfv\'{e}n-Cyclotron Fluctuations: Application
to Solar Wind Turbulence Spectrum}

\author{Yan Wei Jiang,\altaffilmark{1}, Siming Liu,\altaffilmark{2}  Vah\'e
Petrosian,\altaffilmark{1,3}, Christopher L. Fryer,\altaffilmark{2} and Hui 
Li\altaffilmark{2}}

\altaffiltext{1}{Center for Space Science and Astrophysics, Department of Physics, 
Stanford University, Stanford, CA 94305; arjiang@stanford.edu, vahe@astronomy.stanford.edu}
\altaffiltext{2}{Los Alamos National Laboratory, Los Alamos, NM, 87545; 
liusm@lanl.gov, fryer@lanl.gov, hli@lanl.gov}
\altaffiltext{3}{Also Department of Applied Physics}

\begin{abstract}

With the diffusion approximation, we study the cascade and damping of Alfv\'{e}n-cyclotron fluctuations in solar plasmas numerically. Motivated by wave-wave couplings and nonlinear effects, we test several forms of 
the diffusion tensor. For a general locally anisotropic and inhomogeneous diffusion tensor in the wave vector 
space, the turbulence spectrum in the inertial range can be fitted with power-laws with the power-law index 
varying with the wave propagation direction. For several locally isotropic but inhomogeneous diffusion 
coefficients, the steady-state turbulence spectra are nearly isotropic in the absence of damping and can be 
fitted by a single power-law function. However, the energy flux is strongly polarized due to the 
inhomogeneity that leads to an anisotropic cascade. Including the anisotropic thermal damping, the 
turbulence spectrum cuts off at the wave numbers, where the damping rates become comparable to the cascade rates. 
The combined anisotropic effects of cascade and damping make this cutoff wave number dependent on the wave 
propagation direction, and the propagation direction integrated turbulence spectrum resembles a broken 
power-law, which cuts off at the maximum of the cutoff wave numbers or the $^4$He cyclotron frequency. Taking 
into account the Doppler effects, the model can naturally reproduce the broken power-law wave spectra observed 
in the solar wind and predicts that a higher break frequency is aways accompanied with a greater spectral 
index change that may be caused by the increase of the Alfv\'{e}n Mach number, the reciprocal of the 
plasma beta, and/or the angle between the solar wind velocity and the mean magnetic field. These predictions 
can be tested by future observations.


\end{abstract}





\section{Introduction}

Turbulence is ubiquitous in the universe and plays important roles in our understanding of many natural 
phenomena (Kolmogorov 1941; Iroshnikov 1984; Kraichnan 1965). It occurs in highly non-equilibrium systems, 
where the microscopic viscous and/or resistive dissipation processes can not effectively convert the free 
energy into the internal energy of the fluid. Such systems usually have very high Reynold numbers and the free 
energy is stored in the large scale motion and/or magnetic fields. For plasmas, the free energy initially may 
also be stored in non-equilibrium distributions of charged particles. Turbulence is generated through a 
variety of instabilities related either to the large scale (magneto-)hydrodynamical processes or the 
microscopic collective plasma effects or plasma physics processes. These aspects have been extensively 
investigated with (magneto-)hydrodynamics and/or plasma physics theories.

In astrophysics, most turbulence is carried by magnetized plasmas. It is responsible for distributing energies among 
different components of the plasmas, which may result in distinct emission characteristics or other observable 
features. Observations of these radiations can then be used to study the corresponding astrophysical sources 
(e.g. Liu et al. 2004; 2006 on $^3$He rich solar flares). Many of the astrophysical plasmas we are interested in 
are made of charged particles. Depending on the turbulence energy loss rate to these background particles, its 
energy evolution is usually separated into two phases or ranges: cascade and damping. For the former, energy is 
usually transferred from large scales to small scales, and the turbulent motion has weak coupling with the 
background particles so that there are no significant energy exchanges between them. The corresponding spectral 
range is called the inertial range, where the energy flux is independent of the spatial scale. 

In the damping phase, there are strong couplings between the charged background particles and turbulent motion, 
and the background particles are energized. For collisional plasmas, where the Coulomb collision timescales are 
much shorter than other relevant timescales, the background particles very quickly reach a thermal equilibrium. The 
energy partition in such systems is relatively simple, and its studies focus on other aspects, such as radiative 
processes, radiation transfer, ionization, and elements mixing. Much astrophysical turbulence, however, is 
carried by collisionless plasmas, where the Coulomb collision timescales are long. Although it is generally 
accepted that the particle distributions in these plasmas are determined by the coupling of these charged 
particles with the turbulent electro-magnetic fluctuations, the details of these interactions are not well 
understood. Given the high degree of freedom to characteristize the particle distributions, it is usually 
assumed that the background particles are in thermal equilibrium at least with their kinds. The damping 
therefore occurs on the smallest spacial scales and has been an essential part of plasma physics theories for 
these kinds of collisionless thermal plasmas (Andr\'{e} 1985; Gary \& Borovsky 2004). The damping of fast-mode 
waves by isotropic power-law electron and proton populations under typical solar-flare conditions was recently 
studied by Petrosian et al. (2006). Large scale waves can be damped by high energy particles through resonant 
wave-particle couplings in the case.

The cascade has been an essential element in all kinds of turbulence studies. The highly nonlinear nature of 
turbulence makes this a very challenging object. Nevertheless, the energy transfer of isotropic hydrodynamical 
turbulence, the simplest form of turbulence, has been described reasonably well with the Kolmogorov 
phenomenology that assumes a scale-independent self-similar cascade process. This assumption, in combination 
with the fact that turbulence energy is usually injected at large scales and dissipated at small scales, leads 
to the famous Kolmogorov power spectrum with a power-law spectral index of $5/3$ in the inertial range 
(Kolmogorov 1941). Most astrophysical phasmas carry magnetic fields so that large scale magnetohydrodynamical 
(MHD) waves and small scale kinetic plasma fluctuatioins may be excited. The wave propagation effects on the 
cascade of MHD turbulence were first discussed by Iroshnikov (1963) and Kraichnan (1965). Although MHD effects 
are expected to introduce anisotropy to the system that can affect the couplings between the turbulence and 
background particles significantly, these preliminary investigations assumed that the turbulence is isotropic 
and reduced the cascade to a 1-D problem. The reduction of the triple correlation time to the wave period gives 
rise to a power-law spectral index of $3/2$ (Zhou \& Matthaeus 1990). These are called the Iroshnikov-Kraichnan 
(IK) phenomenology. Since the Kolmogorov cascade time scales as $k^{-2/3}$, where $k$ is the wave number, (the IK 
cascade time scales as $k^{-1/2}$) and the periods of MHD waves scale as $k^{-1}$, the wave propagation effects 
become more important at smaller spatial scales and therefore play critical roles in the energization of 
background particles.

To study the anisotropic effects induced by the presence of a large scale magnetic field and their implications 
on the plasma heating and particle acceleration, MHD turbulence in a uniform medium has been treated as an 
ensemble of linear wave modes (e.g. Petrosian \& Liu 2004). Its nonlinear nature is revealed in the wave-wave 
couplings. Although this quasi-linear treatment may not be valid for strong turbulence, it is certainly a good 
approximation when the magnetic field fluctuations are much smaller than the large scale field and the 
wave periods are much shorter than the eddy turnover timescales. Significant insight of properties of 
Alfv\'{e}n and magnetosonic turbulence has been obtained recently through this approach (Sridhar \& Goldreich
1994; Goldreich \& Sridhar 1995; Galtier et al. 2000; Chandran 2005). However, the component of Alfv\'{e}n 
turbulence excitations nearly perpendicular to a large scale magnetic field is dominated by nonlinear effects 
(Montgomery \& Turner 1981). This indicates inherent limitations of treating MHD turbulence as a spectrum of 
waves. As shown by Montgomery and Matthaeus (1995), linear MHD wave modes do not give a complete description
of turbulence excitations and their coupling, and treating MHD turbulence as an ensemble of linear wave modes
may miss some critical nonlinear effects (Ng \& Bhattacharjee 1996).

Based on the dominance of nonlinear or wave-propagation effects, Oughton et al. (2006) recently separated the 
Alfv\'{e}n turbulence into two interacting parts: quasi-two-dimensional and wave-like fluctuations. The 
quasi-two-dimensional component characterizes the nearly perpendicular excitations and may be described with a 
quasi-2-D Kolmogorov phenomenology. The wave-like fluctuations may be described with an IK phenomenology with 
propagation direction dependent wave periods. The cascade of turbulence then also depends on the assumed 
couplings of these two components. It is interesting to note that the two components are separated by the 
critical balance between linear wave periods and nonlinear eddy turnover timescales, and the incompressible 
Alfv\'{e}n turbulence described by Goldreich \& Sridhar (1995) appears to be appropriate for the 
quasi-two-dimensional component that is dominated by nonlinear effects. Due to the suppression of cascade by 
wave propagation effects, Alfv\'{e}n turbulence with wave periods proportional to the parallel component of wave 
vectors cascades preferentially in the direction perpendicular to the large scale magnetic field. The wave-like 
fluctuations need longer time to develop, which may explain the exponential cutoff of the power spectrum in the 
direction parallel to the large scale magnetic field observed in MHD simulations (Cho, Lazarian, \& Vishniac 
2002).

Advances in computational power and numerical algorithms over the past few decades have made numerical 
simulations one of the important tools for quantitative investigations of magnetized turbulence (Shebalin et al. 1983; 
Cho et al. 2002; Cho \& Lazarian 2003). However, turbulence usually covers a huge dynamical range from the
macroscopic scales of turbulence generation to the microscopic dissipation scales. Current simulations have a
dynamical range of a few hundreds to a few thousands and have not been able to give a complete description of
energy flows in magnetized turbulence. These result in limited applications of them in astrophysics. Moreover,
most of these studies are limited to the MHD regime, where the background particles are strongly coupled with
each other and can be treated together as a single fluid. The electron magnetohydrodynamics (EMHD) treats electrons
and ions as two fluids. It is only applicable in a narrow frequency range between the electron and ion gyro-frequencies
where the whistler dispersion relation is valid (Biskamp et al. 1999; Petrosian \& Liu 2004). To address the
heating of background particles by the turbulence, one has to assume that most of the background particles reach
thermal distributions and arbitrarily extrapolate the turbulence spectrum into the dissipation range, where the
MHD and EMHD formulism is 
usually invalid (e.g. Leamon et al. 1999; Tu et~al. 2002; Zhang \& Li 2004; Wu \& Yang 2006). Observations of
solar flares, solar winds and space plasmas, on the other hand, demand a detailed study of processes in the
dissipation range, where background particles with different charge to mass ratios interact with the electro-magnetic
fluctuations quite differently (Petrosian \& Liu 2004; Liu et al. 2004, 2006; Leamon et al. 1998). This results in
complicated wave dispersion relations, namely the dependence of the wave frequency on the wave vector (Andr\'{e} 1985)
and certainly affects the turbulence cascade. The couplings among waves and particles are even more complicated
(Stix 1962; Xie 2004; Saito \& Gary 2007).

The diffusion approximation for the power spectrum in the wave number space has been a very powerful and 
efficient tool to study the turbulence cascade and dissipation over a large dynamical range. The 1-D models not 
only address the transition of turbulence from the large scale Kolmogorov phenomenology to the small scale IK 
cascade (Zhou \& Matthaeus 1990), but also are used to study the acceleration and heating of background 
particles by magnetized turbulence (Miller et al. 1995, 1996) and the damping of waves at small scales (Li et 
al. 2001). These studies have deepened our understanding of energy release processes during solar flares 
significantly. They, however, encounter difficulties in reproducing the broken-power spectrum of solar wind 
magnetic fluctuations (Li et al. 2001; Stawicki et al. 2001). Due to the nonlinear nature of these 1-D diffusion 
models, the turbulence spectrum cuts off sharply at the wave number, where the cascade timescale becomes 
comparable to the wave damping time. The particle acceleration model also needs to be modified to reproduce the 
observed enhancement of high energy $^3$He ions during impulsive solar flares since the 1-D turbulence is damped 
by background $^4$He ions before it reaches the $^3$He gyro-frequency to accelerate the low energy $^3$He ions
from a thermal background (Mason et al. 2002; Liu et al. 2004, 2006).

Studies of weak Alfv\'{e}n turbulence with the quasi-linear treatment of electro-magnetic fluctuations have 
shown that the cascade is anisotropic. It is also well-known that damping rates of different plasma wave modes 
by thermal background particles are very sensitive to the wave propagation directions. These anisotropies are 
critical to study energy dissipation through magnetized turbulence in a collisionless thermal plasma. A 2-D
diffusion model is needed to advance of our understanding of magnetized turbulence and address the difficulties
encountered with the 1-D models. Cranmer and Van Ballegooijen (2003) have showed recently that the heating of
the background particles by Alfv\'{e}n turbulence are very sensitive to the 2-D and kinetic effects. To recover
the critical balance and partially take into account the kinetic effects, they constructed a complicated
diffusion-convection equation for the power spectrum with three dimensionless coefficients. A similar
quasi-2-D model was proposed recently by Howes et al. (2007) to explain the broken power-law power spectrum of the solar wind turbulence.

In this paper, we study the general characteristics of the nonlinear 2-D diffusion model, the kinetic and damping 
effects of the Alfv\'{e}n-cyclotron fluctuations. In principle, all plasma wave modes, such as Alfv\'{e}n, fast 
and slow modes, should be included to have a complete description of the cascade and damping of magnetized 
turbulence. One then needs to solve a set of coupled diffusion equations for each wave branches (Andr\'{e} 
1985). This is a quite challenging task since the results will depend both on the interactions within each wave 
branch and couplings among different branches. Fortunately, Cho and Lazarian (2003) have shown that couplings
among different wave branches are usually weak and these couplings decrease toward small spatial scales (Luo \& 
Melrose 2006). Chandran (2005) showed that the couplings between Alfv\'{e}n and fast modes are significant only 
in the direction parallel to the large scale magnetic field, where the frequencies of the Alfv\'{e}n and fast
modes are comparable. The couplings among different branches therefore can be separated from other processes near 
the dissipation range, where the kinetic and damping effects dominate. We use the exact dispersion relation
for a cold plasma, which is a good approximation for the more general dispersion relation of a collisionless
thermal plasma. The diffusion coefficients are constructed as a function of the wave phase and/or group
velocities so that the kinetic effects are treated self-consistently. The linear Vlasov equation is used to
derive the thermal wave damping rates. Since the damping rate increases sharply with the increase of wave number.
This quasi-linear treatment is expected to give a good approximation of the damping even for the nonlinear effect
dominated quasi-two-dimentional component. The diffusion model actually does not distinguish the quasi-two-dimentional
and wave-like fluctuations (Oughton et al. 2006). The balance between eddy turnover and wave propagation is
revealed in the diffusion tensor.

The Alfv\'{e}n-cyclotron branch has been studied extensively due to its simplicity and its prevalence in
magnetized turbulence. It is chosen here to facilitate better comparisons of our model with previous results.
In \S\ \ref{general}, we discuss how the turbulence cascade can be studied using the diffusion approximation,
which reduces the turbulence evolution to a nonlinear 2-D diffusion problem. The nonlinear diffusion equation
can be solved numerically to obtain the power spectrum from the MHD region to the $^4$He gyro-frequency,
where the Alfv\'{e}n dispersion surface cuts off. To compare with previous studies, we first presents the
results for the cascade of Alfv\'{e}n turbulence \S\ \ref{cascade}, where the dispersion relation for Alfv\'{e}n
waves is adopted. The exact dispersion relation and damping rate are discussed in \S\ \ref{disp}. The kinetic
and damping effects are investigated in \S\ \ref{result}, where we also consider its application to solar
wind magnetic fluctuations. We discuss the implication of this model on the study of plasma heating and
particle acceleration by magnetized turbulence, future work, and draw conclusion in \S\ \ref{dis}.

\section{Kinetic Equation and Diffusion Tensor}
\label{general}

For weak or intermediate level turbulence, the most efficient way of energy cascading is through 3-wave 
interactions, with the resonance condition
\begin{eqnarray}
{\bf k_1}+{\bf k_2}={\bf k}, \quad
\omega({\bf k_1})+\omega({\bf k_2})=\omega({\bf k}),
\label{reseq}
\end{eqnarray}
where ${\bf k}$ and $\omega$ are the wave vector and the wave frequency, respectively, and  $\omega({\bf k})$ gives 
the wave dispersion relation. For Alfv\'{e}n waves, $\omega({\bf k})= v_{\rm A} k_{||}$, where $v_{\rm A}$ is the 
Alfv\'{e}n velocity and $k_{||}$ is the parallel component of the wave vector. Since only oppositely directed wave 
packets can interact, this resonance condition further requires one of the interacting wave vectors, say ${\bf k_2}$, 
has to be perpendicular to the large scale magnetic, i.e., $k_{2||}=0$ (Shebalin, Matthaeus \& Montgomery 1983). As a 
result, Alfv\'{e}n wave turbulence cascades strongly in the direction perpendicular to the mean magnetic field. In 
the parallel direction, hydrodynamic eddy interactions prevent the wave cascading from vanishing but this can be a 
slow process. 

To study the wave cascade through the above 3-wave interactions, Goldreich \& Shridhar (1995) first wrote 
down the integral form of the kinetic equation for the wave power spectrum ${\cal W}({\bf k}, t)$
\begin{equation}
{\partial {\cal W}({\bf k}, t) \over \partial t}=4\int d^3k_1 d^3k_2 \delta({\bf k_1}+{\bf k_2}-{\bf k})
	(\bf{\hat{e}_1} \cdot \bf{\hat{e}}_2)^2({\bf k} \cdot \bf{\hat{e}}_2)^2
	{\cal W}({\bf k_2})[{\cal W}({\bf k_1})-{\cal W}({\bf k})]{\cal T}({\bf k_2})\,,
\label{GSInt}
\end{equation}
where  $\bf{\hat{e}_j}$ ($j=1,2$) are the unit polarization vector of the specific Alfv\'{e}n waves. ${\cal 
T}({\bf k})={\eta({\bf k})/[\eta({\bf k})^2+4\omega({\bf k})^2]}$ is the timescale of wave-wave interaction 
determined by the ``eddy damping rate" $\eta$ and the wave frequency. This equation strictly satisfies the 3-wave 
resonance condition and only allows energy cascade in the perpendicular direction. It does not take into account the 
nonlinear effects of turbulence and therefore cannot describe the general turbulence cascading in the 3-D wave vector 
space. By including Alfv\'{e}n-fast mode cross interaction, Chandran (2005) found that waves can cascade in all 
directions and solved the coupled kinetic equations for the Alfv\'{e}n and fast mode wave power spectra 
simultaneously. However, the Alfv\'{e}n wave spectrum does not appear to have any feature associated with 
``critical balance'' first introduced by Goldreich \& Shridhar (1994, 1995). 

In this paper, we focus on Alfv\'{e}n-Alfv\'{e}n wave interactions and extend the theory beyond the MHD regime by 
using the general dispersion relation for the Alfv\'{e}n-cyclotron fluctuations. To incorporate the 
nonlinear effects caused by the eddy turnovers, we adopt the diffusion approximation, which leads to the general 
kinetic equation:
\begin{eqnarray}
   {\partial {\cal W}({\bf k}, t) \over \partial t}
& = &\dot{Q}_{\cal W}({\bf k}, t) 
+ {\partial\over\partial k_i}\left[D_{ij}{\partial\over \partial k_j}{\cal W}({\bf k}, t)\right] 
- \Gamma({\bf k}){\cal W}({\bf k}, t)
- {{\cal W}({\bf k}, t) \over T^{\cal W}_{\rm esc}({\bf k})}\,,
\label{WHOMOG}
\end{eqnarray}
where the terms on the right-hand-side represent the wave generation, cascade, damping and leakage processes, 
respectively. In the following, we will ignore the leakage term and treat the source term as a $\delta$-function at 
certain large scale and study how the wave spectrum depends on the cascading and damping processes. 
By proper choices of the diffusion tensor, the cascade caused by the wave-wave resonances and eddy 
turnovers may be modeled (Zhou \& Matthaeus 1990). We are interested in the solution that decreases rapidly with 
the increase of the wave number. Equation (\ref{GSInt}) then shows that contributions of the 3-wave interactions 
to the evolution of the power spectrum is dominated by processes with $k_2\ll 1$. This justifies the diffusion 
approximation for the 3-wave interactions. For $k_2\ll 1$, $({\cal W}_1-{\cal W}_k) \simeq {\bf k_2}\cdot\nabla 
{\cal W}({\bf k})$, which corresponds to a flux in the wave vector space. Comparing equations (\ref{GSInt}) and 
(\ref{WHOMOG}), it is obvious that the amplitude of the diffusion tensor is proportional to the total turbulence 
energy.
   
Following Zhou \& Matthaeus's (1990) theory, we construct the diffusion tensor with the nonlinear timescale 
$\tau_{\rm NL}$ associated with eddy turnovers and wave crossing time $\tau_{\rm A}$. 
The eddy turnover time can be estimated with 
$\tau_{\rm NL}=(v k)^{-1}$, where the eddy velocity $v\simeq ({\cal W} k^3)^{1/2}$. For the wave crossing time, 
one must consider the anisotropy of Alfv\'{e}n-cyclotron dispersion. Near the perpendicular direction, since 
$k_{2,||}\simeq 0$, any wave can always interact with this large scale wave, i.e. $\tau_{\rm A}\sim \infty$. The 
wave crossing effect can be ignored, and the cascade is dominated by the eddy turnover; In the other directions, 
wave packets with a size of $\sim 1/k_{||}$ cross each other at the Alfv\'{e}n speed. In general, we have 
$\tau_{\rm A}\simeq(v_{\rm A} k_{||})^{-1}.$\footnote{Note that, following Kraichnan's (1965) argument for an 
isotropic Alfv\'{e}n wave turbulence, Zhou \& Matthaeus (1990) obtained $\tau_A=(v_A k)^{-1}$ for their 1-D diffusion 
model.} Then the 3-wave coupling time $\tau_3=(\tau_{\rm NL}^{-1}+\tau_{\rm W}^{-1})^{-1}$, and the wave 
cascading rate $\tau_{\rm cas}^{-1}\simeq\tau_{\rm NL}^{-2}\tau_3$. Then the locally isotropic diffusion tensor
\begin{equation}
D_{ij} 
=\delta_{ij} {C \over 4\pi} k^2 \tau_{\rm cas}^{-1}
=\delta_{ij}{C \over 4\pi}{{\cal W} k^7 \over ({\cal W} k^3)^{1/2} k + v_A k \cos\theta}
\label{DijMHD}
\end{equation}
where $\delta_{ij}$ is Kronecker's delta with $i, j$ indicating the three bases of the 3-D wave vector space,
$C$ is a dimentionless cascading constant corresponding to the Kolmogoroff constant for hydrodynamic turbulence, 
and $\theta$ is the angle between the wave vector ${\bf k}$ and the mean magnetic field.

\section{Cascade of Alfv\'{e}n Turbulence}

\label{cascade}

\begin{figure}[thb]
\begin{center}
\includegraphics[width=0.45\textwidth]{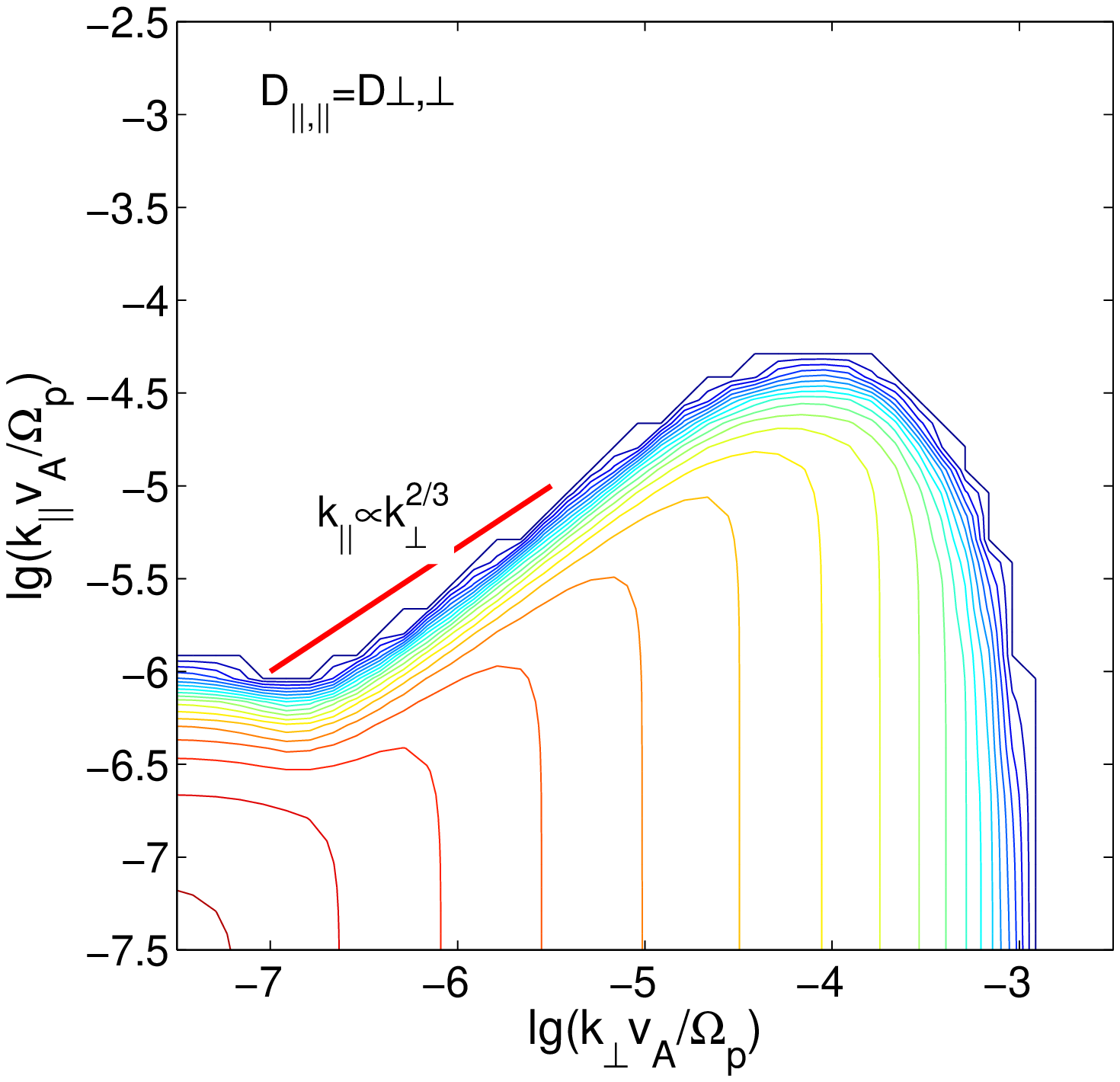}
\includegraphics[width=0.45\textwidth]{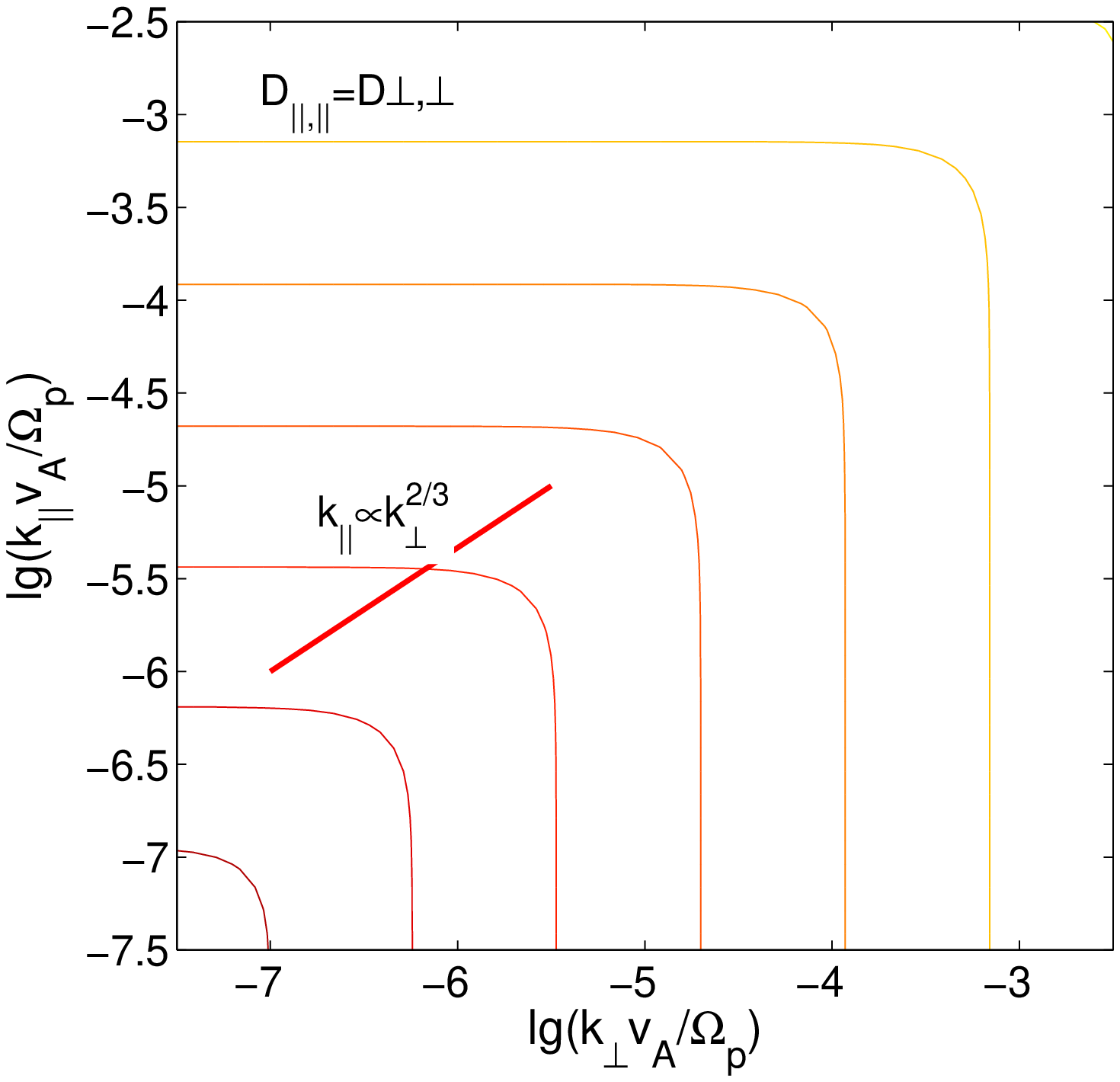}
\end{center}
\begin{center}
\includegraphics[width=0.45\textwidth]{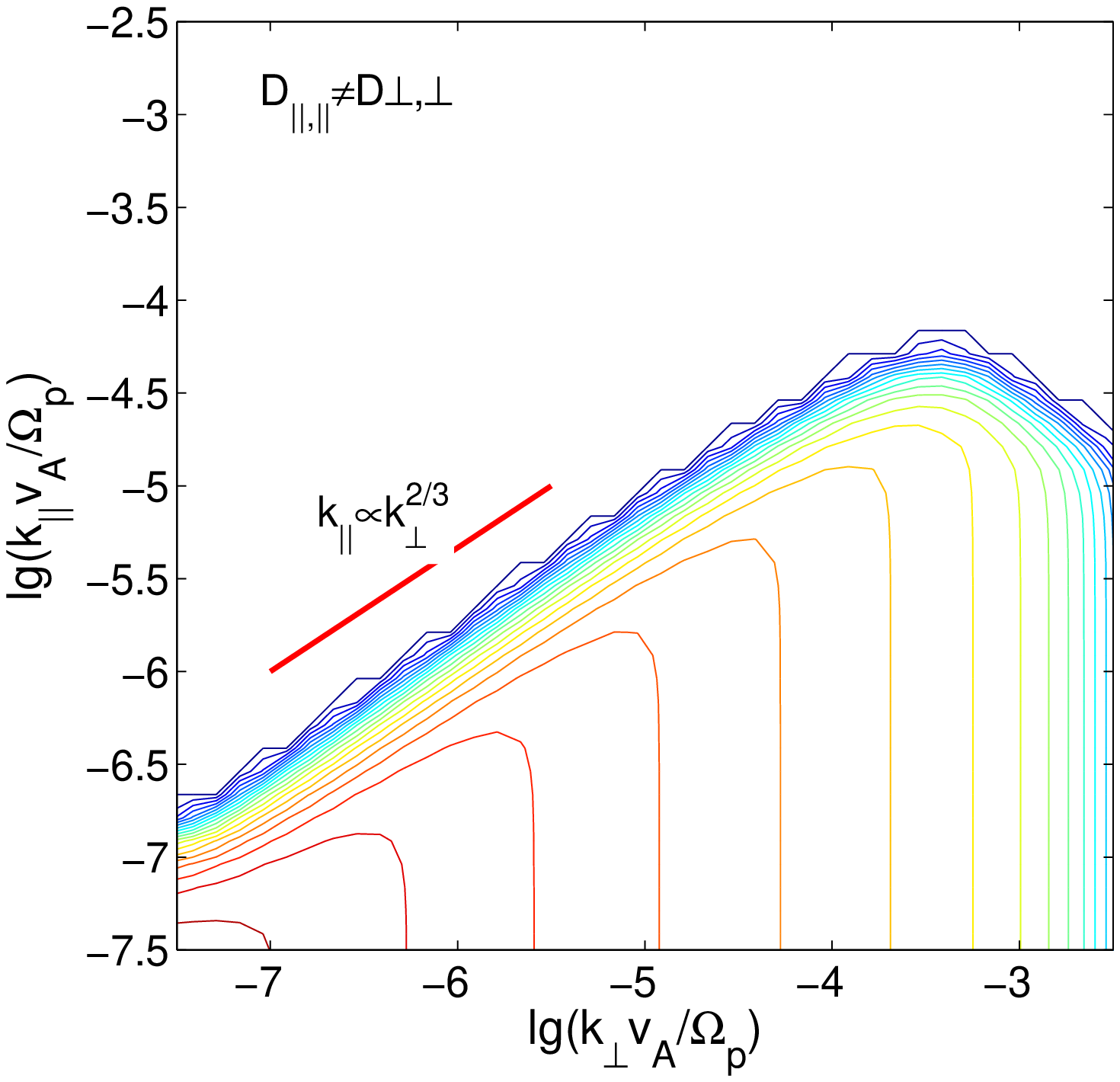}
\includegraphics[width=0.45\textwidth]{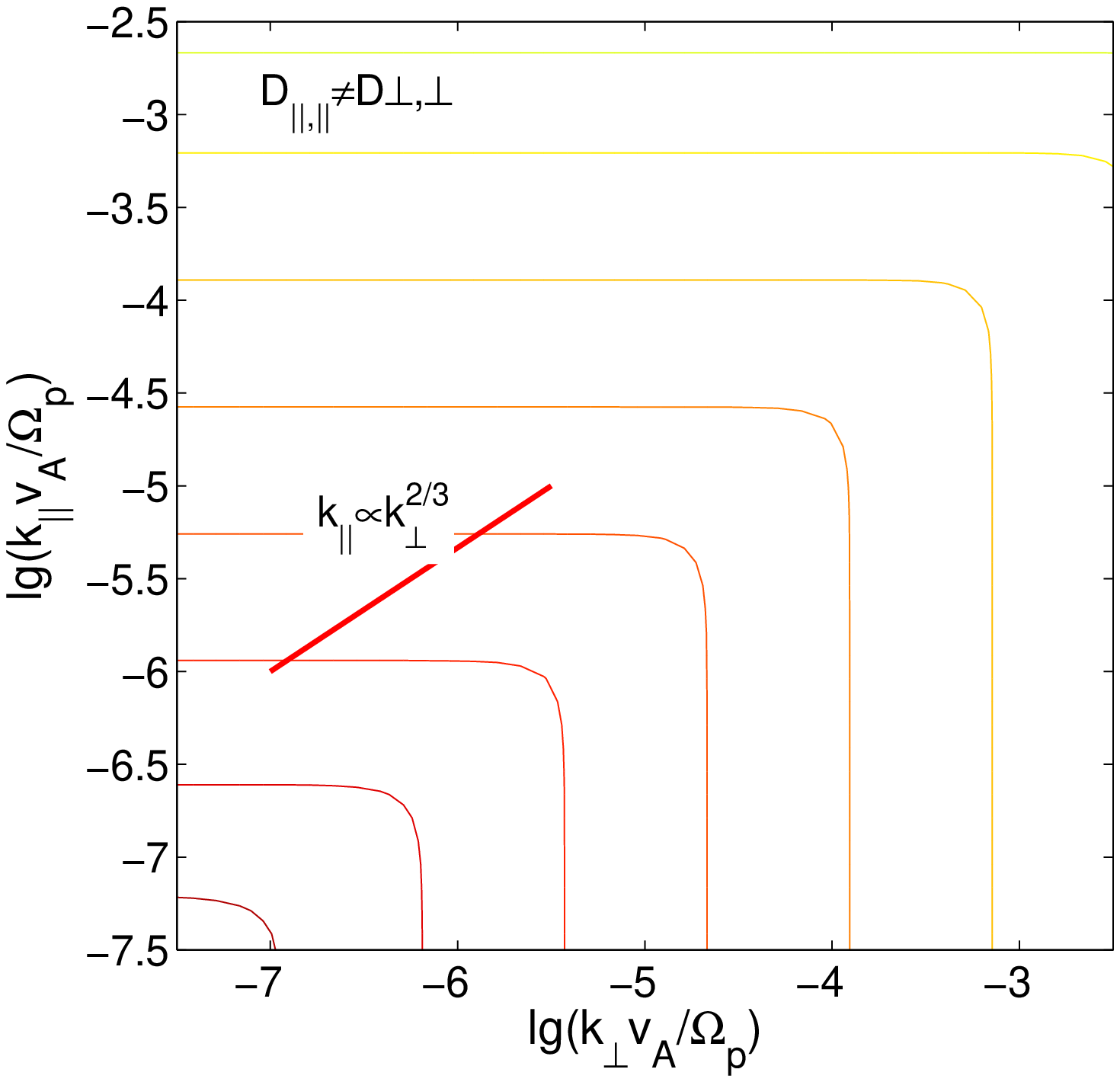}
\end{center}
\caption{
Wave energy ${\cal W}(k_{||},k_{\perp})$ contours at intermediate (left panels) and 
steady state (right panels) for weak turbulence without damping. The upper panels show 
the result of isotropic model ($D_{||,||}=D_{\perp,\perp}$) and lower panels show the
result of anisotropic model $D_{||,||}\neq D_{\perp,\perp}$. The red line labels the
critical balance relation introduced by Sridhar \& Goldreich (1994).
}
\label{Acon.ps}
\end{figure}

In the strong turbulence limit, $v \gg v_A$, this diffusion tensor recovers the isotropic Kolmogoroff cascade. 
However, in the weak turbulence limit, i.e. $v \ll v_A$, it does not reproduce the isotropic
Kraichnan spectrum due to the global anisotropy of the diffusion coefficient caused by the anisotropic Alfv\'{e}n 
wave crossing time. Since the Alfv\'{e}n wave crossing time $\tau_A$ is much longer in the perpendicular 
direction than in the parallel direction, the wave cascading rate is large in perpendicular direction and 
decreases dramatically with the increase of $k_{||}$. The wave energy contours in Figure \ref{Acon.ps} show the 
numerical result of such anisotropic cascading. The turbulence energy cuts off at the cascading front. In fact, 
if we assume a Kolmogoroff power spectrum, the Alfven wave crossing process starts to dominate and suppress the 
wave cascading at $k_{||} \gtrsim k_{\perp}^{2/3}$, which is similar to the critical balance relation introduced 
by Goldreich \& Shridhar and explains the $k_{||} \sim k_{\perp}^{2/3}$ scaling of the cascading front. 
This cascading front in wave vector space extends until it reaches the numerical or physical boundary 
(damping or non-MHD effect at large $k$). However, although the cascading energy flux is highly 
concentrated on the perpendicular direction at small $k_{||}$, the diffusive process still carries wave
energy to large $k_{||}$. By assuming an infinite damping at certain 
large $k$ and let turbulence evolve freely at any $k$ below this 
infinite damping boundary, we can see (Figure \ref{Acon.ps}) the turbulence fills all possible wave 
vector space in our numerical simulation. In our model, the energy transfer to large $k_{||}$
through two mechanisms: the reverse cascading on perpendicular 
direction and the possible slow cascading in parallel direction. For MHD turbulence, 
many publications () consider the cascading in parallel direction void beyond certain critical value, 
i.e. $k_{||} \sim k_{\perp}^{2/3}$. However, there is no theory against inverse cascading on 
perpendicular direction. As a result, we argue that our diffusion theory is valid and the wave energy 
fills all possible wave vectors before it gets damped. 

\begin{figure}[thb]
\epsscale{0.6}
\centerline{\plotone{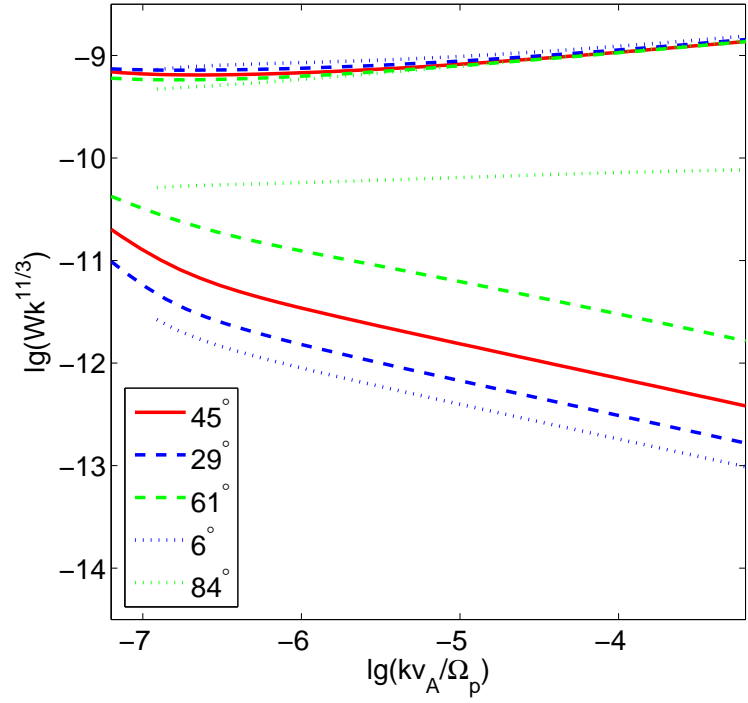}}
\caption{
Power spectra of Alfv\'{e}n wave turbulence for two diffusion tensor models. The result from
isotropic model ($D_{||,||}=D_{\perp,\perp}$) is shifted by 10 times. The isotropic model provides a close to 
isotropic Kolmogoroff spectrum and the anisotropic model $D_{||,||}\neq D_{\perp,\perp}$ provides a steep 
spectra in directions other than perpendicular direction.
}
\label{MHD_ani.ps}
\end{figure}

The ``steady state" spectrum we obtained based on this simple diffusion model is a 
quasi-isotropic Kolmogoroff-like spectrum, with highly anisotropic energy flux concentrated on
perpendicular direction. This is due to an ``isotropic" diffusion tensor (see Figure \ref{Acon.ps}).
Theoretically, by setting the wave cascading rate to be large at small $k_{||}$ and vice versa,
we are able to simulate anisotropic turbulence energy flux concentrated on perpendicular
direction. However, since we set the diffusion coefficient to be the same on parallel and 
perpendicular direction ($D_{||,||}=D_{\perp,\perp}$) for at any ${\bf k}$, we can only reach 
an isotropic spectrum given enough turbulence evolving time. An apparent improvement of this model
is to construct the parallel and perpendicular terms of diffusion tensor differently. We can take
parallel and perpendicular direction as the diffusion tensor's principle axes (eigenvectors) and
the corresponding diffusion coefficients as its eigenvalues.
For cascading on perpendicular direction, since $k_{2,||}=0$, the wave cascading is limited by wave eddy turn 
over time only, then we can write the $D_{\perp,\perp}=k^2\tau^{-1}_{cas,\perp}=k^2 \tau^{-1}_{NL}$.
For cascading on parallel direction, small scale wave $\omega({\bf k_1})$ collides with large 
scale wave $\omega({\bf k_2})$ and gradually obtain a 
drift in wavelength (${\bf k_1}\to{\bf k}$). Although this drift can be both way (to smaller 
or larger scales), the 
energy gradient in wave vector space keeps the energy flow from large to small scales stronger than the 
inverse. Since we are interested 
in the change of wavelength during a single wave period of  ${\bf k_1}$, its own frequency $\omega({\bf k_1})$
limits the cascading rate. Thus we combine this wave frequency with eddy turn over rate to obtain cascading rate
on parallel direction $\tau^{-1}_{cas,||}=\tau^{-2}_{NL}/(\tau_{NL}^{-1}+\tau_A^{-1})$. And the 
corresponding diffusion tensor can be written as $D_{||,||}=k^2\tau^{-1}_{cas,||}$. Thus,
we obtain the form of the anisotropic diffusion tensor as
\begin{equation}
D_{ij} = {C \over 4\pi} k^2 \left(
\begin{array}{cc}
{\tau_{NL}^{-2} \over \tau_{NL}^{-1}+\tau_A^{-1}} & 0 \\
0 & \tau_{NL}^{-1}\\
\end{array}
\right),
\label{Dij_ani.eq}
\end{equation}
where $\tau_{NL}^{-1}$ and $\tau_A^{-1}$ is defined the same way as in isotropic diffusion model. The 
resulting spectra in different directions (see Figure \ref{MHD_ani.ps}) show such an anisotropy. Note
that for both isotropic and non-isotropic 
diffusion tensor model, we have larger cascading rate at small $k_{||}$, and the energy flux is highly 
concentrated on perpendicular direction.

Instead of exploring more possibilities on the forms of diffusion tensor, we refer to observation of turbulence spectrum
in solar wind to point us the direction. The discussion can be found in \S\ \ref{result}. 

\section{Dispersion Surface beyond MHD Region and Thermal Damping}
\label{disp}

The Solar Wind observation suggests that the Alfv\'{e}n wave turbulence cascade to smaller scales than the
MHD regime. Along parallel direction (for all $k_{\perp}<1/\rho_p$, where $\rho_p$ is proton 
Lamour radius), the Alfven wave dispersion surface diverge from simple 
$\omega \approx k v_A\cos\theta$ around $k v_A\cos\theta \lesssim 0.5 \Omega_p$ and flatten into He cyclotron
oscillation ($\omega = 0.5 \Omega_p$) for larger $k_{||}$s. We discuss how this affects the wave energy cascading 
below. On the perpendicular direction, the dispersion surface bends
into kinetic Alfv\'{e}n wave ({\bf KAW}) at $k \rho_p \sim 1$. The effects of KAW dispersion relation 
and its high damping rate on turbulence cascading is even more complex. However, 
as we will discuss in \S\ \ref{result}, in typical solar wind conditions (where plasma beta $\beta_p<1$
or $\sim 1$), KAW is Doppler-shifted to a high frequency around or beyond the satellite 
observation limit. Thereafter, we limit our study to a wave vector space below KAW range in this paper.

When the damping rate is small, the hot plasma dispersion surface can be calculated through numerically 
solving Vlasov equation, where Lauren series is used to approximate the Z-function in the solution. 
Because of this approximation, when the damping rate is getting close to the wave
frequency from below, this numerical method fails and there is no simple scheme that can 
guarantee a solution. Figure \ref{disps.ps} shows the numerical result of Alfv\'{e}n wave 
dispersion surface using Waves in Homogeneous, Anisotropic, Multicomponent Plasmas ({\bf WHAMP}) code 
(Ronnmark 1982). The missing segments in the dispersion relation are the wave vectors 
where the numerical solution fails due to the relative large damping rate. 
Fortunately, as we can also see from Figure 5, the cold plasma approximation provides a
close fit to the ``real" dispersion relation in the wave vector range of our interest (i.e. beyond MHD but below KAW). 
Since we have the analytical form of cold plasma dispersion relation (see Appendix \ref{cold}), 
we can solve the smooth function wave frequency $\omega({\bf k})$ and its gradient 
(i.e. wave group velocity)
$\bf{v_g}=\bigtriangledown_{\bf{k}}\omega(\bf{k})$ at all the wave vector grid points for the use of
our numerical simulation on turbulence energy cascading. Thereafter, we use cold plasma dispersion 
relation to study this diffusion process in the following.

\begin{figure}[thb]
\epsscale{0.6}
\centerline{\plotone{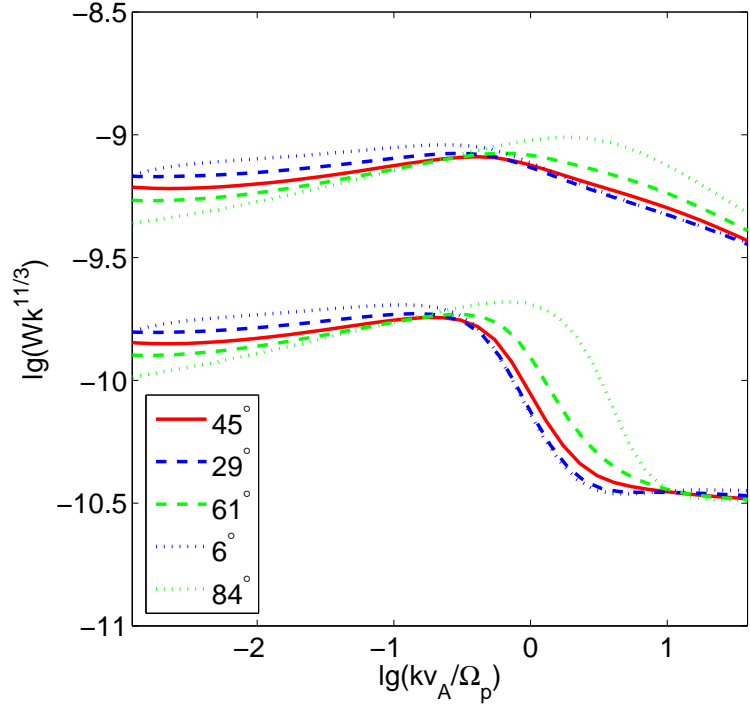}}
\caption{
Wave energy spectra with two different wave crossing rates beyond MHD regime.
The breaks at $k_{||} v_A \sim 0.5 \Omega_p$ in both spectra are due to the 
dispersion relation $\omega({\bf k})$ deviating from the simple MHD relation 
$\omega = k_{||} v_A$. 
The two spectra, as discussed later, can not be differentiated by observation 
because of the strong thermal damping for solar wind conditions. 
}
\label{ND.ps}
\end{figure}

\begin{figure}[htb]
\begin{center}
\centerline{\plotone{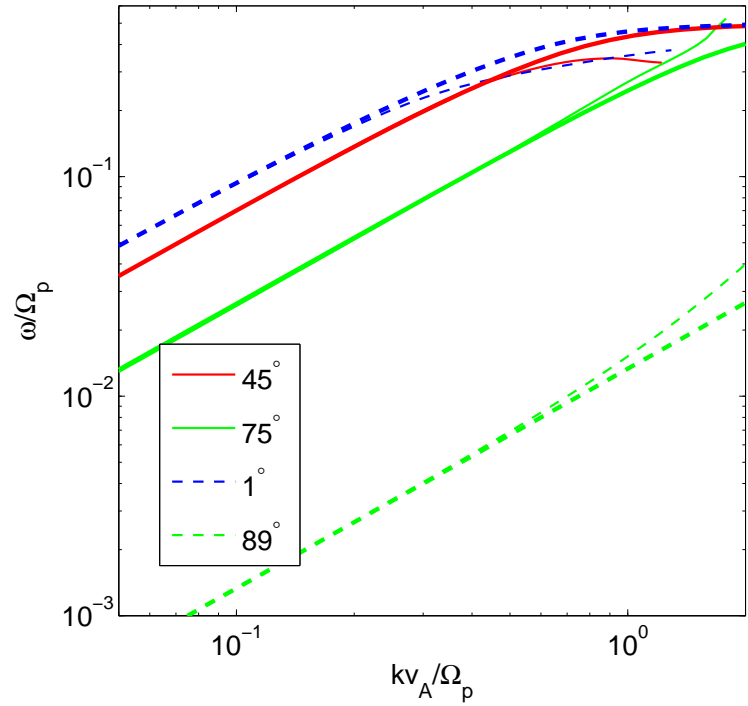}}
\end{center}
\caption{
Plasma dispersion relations under typical solar wind conditions comparing the one under cold plasma approximation.
The former one is solved numerically with WHAMP code.
}
\label{disps.ps} 
\end{figure}

It requires a detailed study to understand how the dispersion surface beyond the MHD regime 
affects wave cascading. In the region where the dispersion surface starts to diverge from MHD
approximation, i.e. $k_{||} v_A \sim 0.5 \Omega_p$, this surface, 
like in MHD regime, is flat on perpendicular direction 
(i.e. $\partial \omega /\partial k_{\perp}\ll \partial \omega /\partial k_{||}$). It's easy to 
demonstrate that 
the 3-wave resonance condition (Equation (\ref{reseq})) still requires the smaller one of the interacting
wave vectors ${\bf k_2}$ to be in perpendicular or near perpendicular direction.
So, at region $k_{||} v_A \sim 0.5 \Omega_{p}$, the Alfv\'{e}n 
wave cascading is still highly preferred in perpendicular direction and we can simply adapt the 
wave crossing time as $\tau_A={\bf v_g \cdot k} \simeq \partial \omega /\partial k_{||} k$ in our diffusion 
tensor. In the region far beyond MHD,
i.e. $k_{||} v_A \gg 0.5 \Omega_{p}$, the cold plasma dispersion surface flattens out in all directions and 
the Alfv\'{e}n wave packet becomes stationary. It is not obvious how the turbulence will cascade in this region. In
one aspect, three wave resonance condition still requires one wave vector to be perpendicular, 
i.e. we need the cascading rate (or diffusion tensor) to be larger in perpendicular direction. 
However, ${\bf v_g \cdot k}=0$ can neither give us such a parallel cascading rate nor the wave crossing time, 
instead $\tau_A=1/\omega({\bf k})$ provides a better approximation.
In the other aspect, turbulence cascading can be as simple hydrodynamic when the oscillation is stationary, 
with eddy turn over rate
serve as the only limit on
wave cascading. Then $\tau_A={\bf v_g \cdot k}=0$ is just what we need 
for the diffusion tensor.
Thereafter we try two different approaches to approximate the wave crossing time: 
$\tau_A=(\bf v_g \cdot {\bf k})^{-1}$ or $\tau_A=\omega^{-1}$. 
The final spectra based these two different wave crossing time are shown in \ref{ND.ps}. They both contain 
a spectral break at $k_{||} v_A \sim \Omega_{\alpha}$, where the wave dispersion surface deviats 
from the MHD approximation. These breaks, though appealing, only have an index change less than one. 
Thereafter neither of these breaks can explain the observed broken power-law spectrum in 
solar wind for Alfv\'{e}n turbulence (Leamon et al. 1989). Furthermore, as we show in \S\ \ref{result}, 
these two wave crossing rate models can can hardly produce any observational difference in 
in solar wind when thermal damping is taken into consideration. Thereafter, we leave this theoretical 
work to further study and use $\tau_A=\omega^{-1}$ in our simulation.

To summarize, we get the simple isotropic form of diffusion tensor $D_{ij}$ for MHD and beyond:
\begin{equation}
D_{ij} =\delta_{ij} {C \over 4\pi} k^2 {\tau_{NL}^{-2} \over \tau_{NL}^{-1}+\tau_A^{-1}}
=\delta_{ij} {C \over 4\pi} {{\cal W} k^7 \over ({\cal W} k^3)^{1/2} k + \omega({\bf k})}
\end{equation}
The form of anisotropic diffusion tensor remains the same as Equation (\ref{Dij_ani.eq}), only that
$\tau_A=\omega^{-1}$.


The wave damping rate, can only be obtained through solving Vlasov equation.
For parallel and quasi-parallel propagating waves, Swanson (1982) simplified the Dielectric Tensor and obtained 
the damping rate for weak damping approximation by electron cyclotron oscillations. 
To study all the damping effects we generalize the formular to calculate the cyclotron damping from
all species of particles:
\begin{displaymath}
\frac{\omega_i}{\omega_r}=-\frac
{\displaystyle\sum_{s}^{} \frac{\sqrt{\pi}\omega^2_{p,s}}{\omega_r k v_s}
exp\biggl[-\left(\frac{\omega-\epsilon_s\Omega_s}{k v_s}\right)^2 \biggr]}
{\displaystyle\frac{2k^2}{\omega^2}+
\displaystyle\sum_{s}^{}\frac{2\omega-\epsilon_s\Omega_s}{\omega(\omega-\epsilon_s\Omega_s)^2}\omega_{p,s}}
\end{displaymath}
where different $s$ stands for different particles. In our study, $s$ ranges from 1 to 3 standing for 
electron, proton and $^4$He respectively.
Note that although this formula gives the value of $\omega_i$ all or $\bf{k}$, the approximation fails at 
$\omega_i\sim\Omega_{c,s}$. At these region when approximation fails, Swanson (1982) shows that $\omega_i \gg \omega_r$.
This result is good enough for our study of wave cascading and damping, because the exact value of damping
rate is not needed for strong damping.

For wave propagating in other angles than parallel, we use WHAMP code to solve the damping 
rate numerically. Similarly, when $\omega_i \sim \omega_r$, the weak turbulence approximation 
is not valid and the code fails to converge. At this region, we roughly estimate the damping rate
with power-law extrapolation (linear in our log-log graph) since the exact value of 
damping is not necessary when it is strong enough to cut off the wave.

\section{Numerical Results and Observation}
\label{result}

We solve the 2-D time dependent wave kinetic equation (\ref{WHOMOG}) with 
Alternative Direction Implicit (ADI) scheme on a log-log uniform grid to study this wave cascading process.
A reflective boundary condition is used at large scale boundary $k_{min}$. For small scale boundary, 
we set $k_{max}$ large enough that all wave energy is damped to zero. We put a small constant 
injection at large scale starting from $t=0$ to simulate the turbulence excitation, i.e. 
$\dot{Q}_{\cal W}({\bf k}, t)=Q_0\delta({\bf k}-{\bf k}_0)\theta(t)$ in wave kinetic equation (Equation (\ref{WHOMOG})).

\subsection{Anisotropy by Isotropic Diffusion tensor}
\label{isoresult}

\begin{figure}[thb]
\begin{center}
\includegraphics[width=0.45\textwidth]{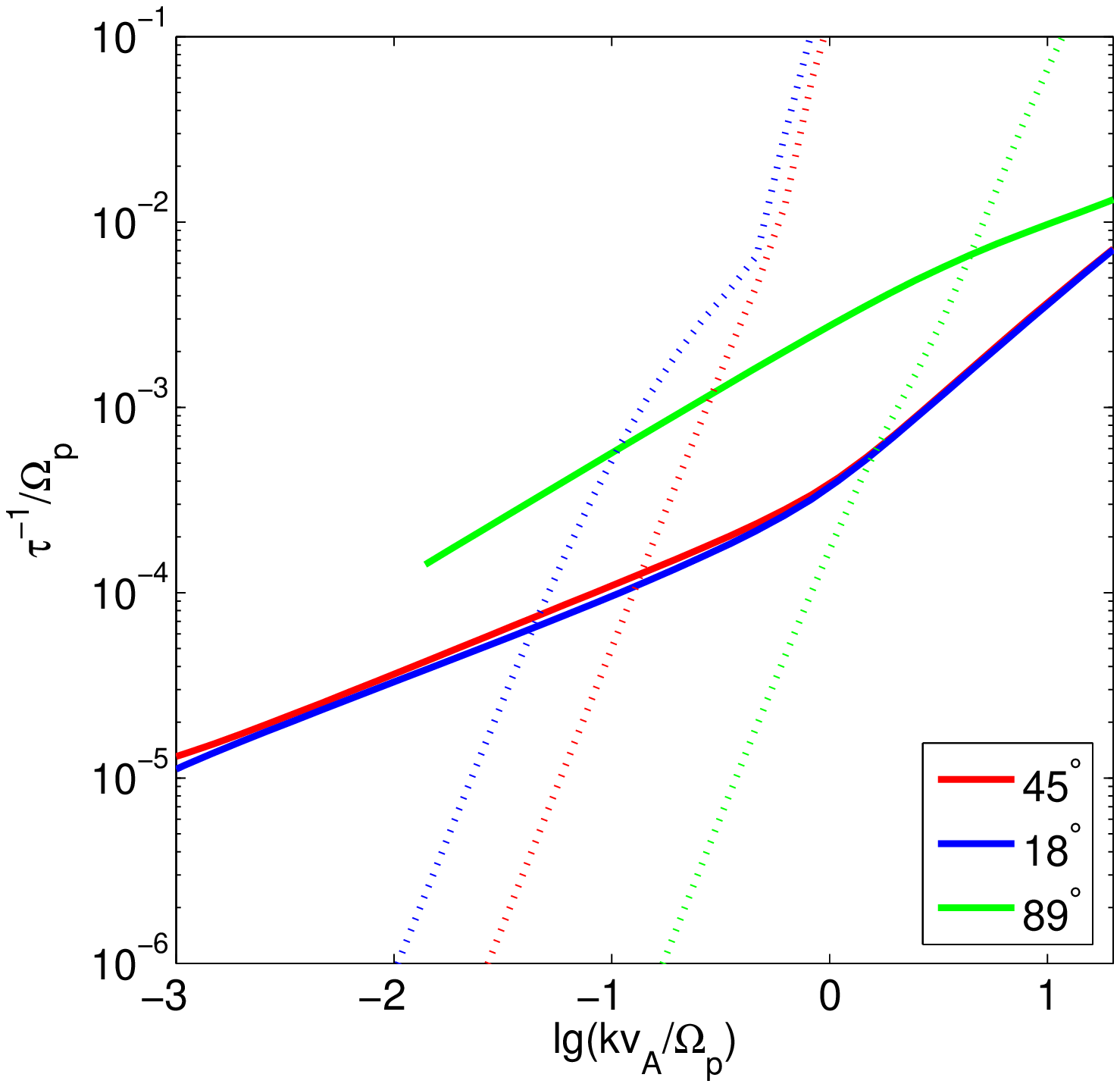}
\includegraphics[width=0.45\textwidth]{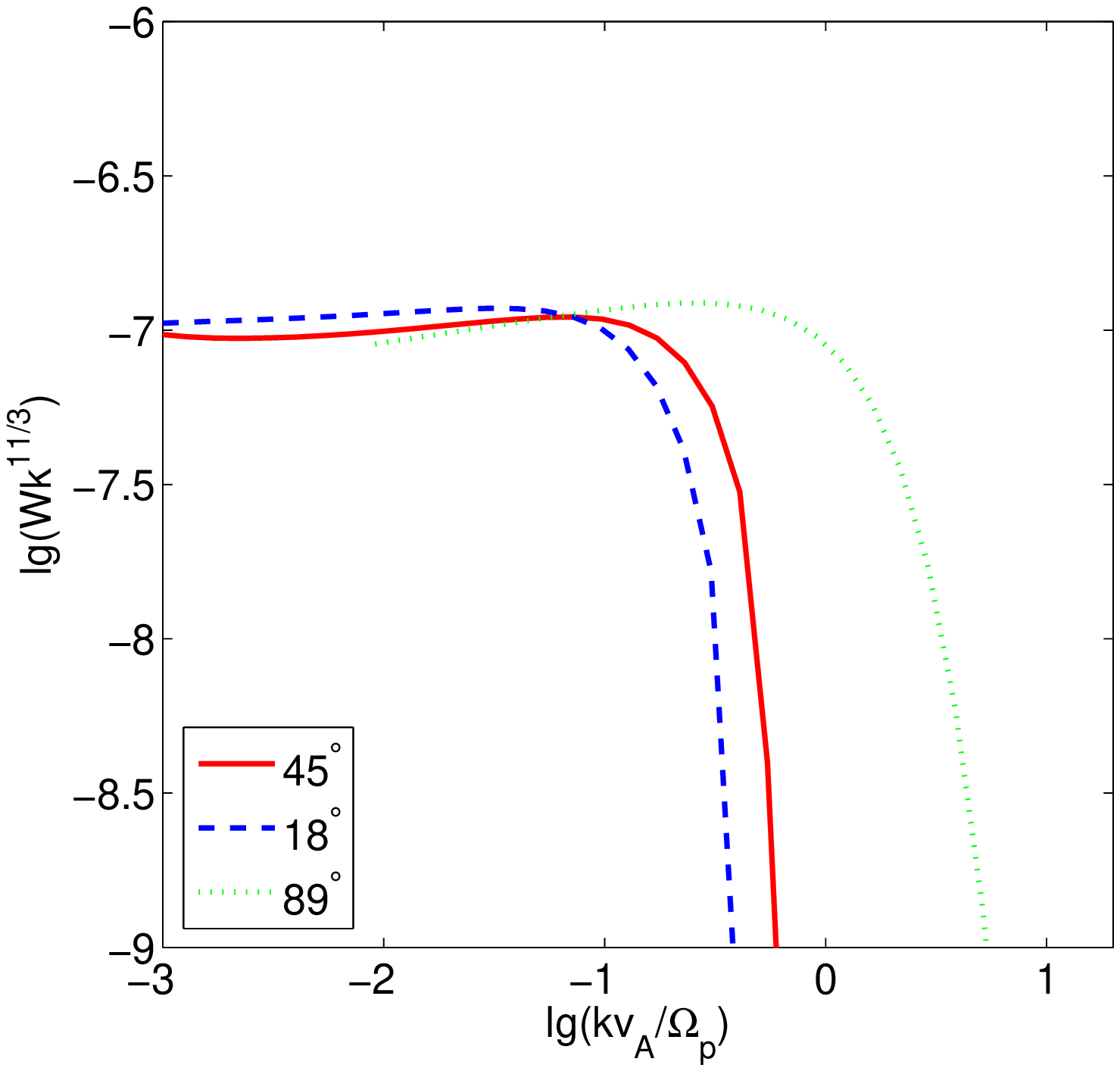}
\end{center}
\caption{
{\it Left}: Wave cascading rate ({\it solid}) vs. damping rate ({\it dotted}) in different direction (with different colors). 
In any direction, the crossing point of two rates gives a rough estimation of where the energy spectrum cuts off. 
{\it Right}: The resulting spectral cut-off that spreads into 1.5 order of magnitude in different direction, 
which can produce a broken power law spectrum when integrated over $\theta$.
}
\label{taus_cut.ps}
\end{figure}

\begin{figure}[htb]
\begin{center}
\includegraphics[width=0.45\textwidth]{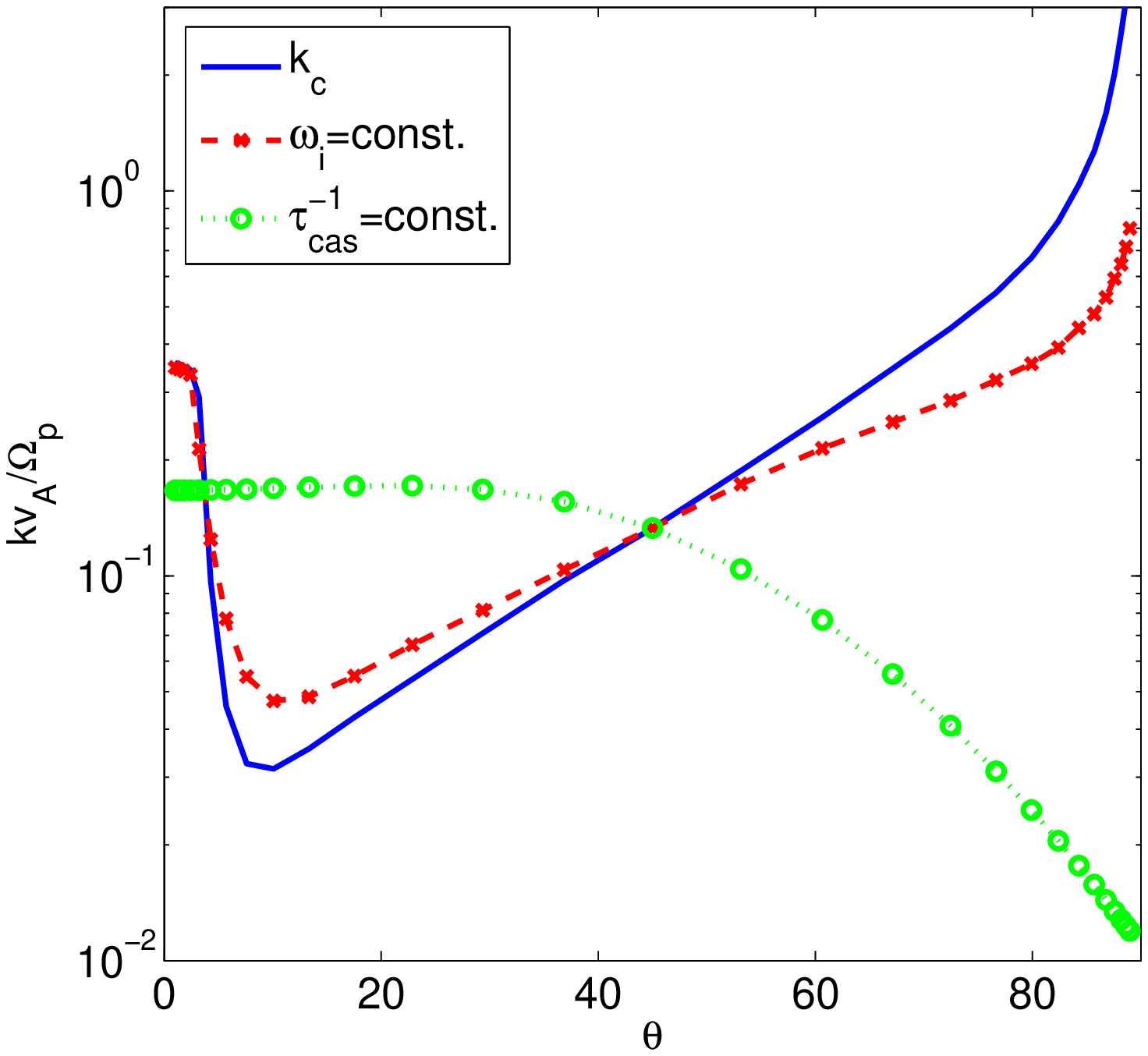}
\includegraphics[width=0.45\textwidth]{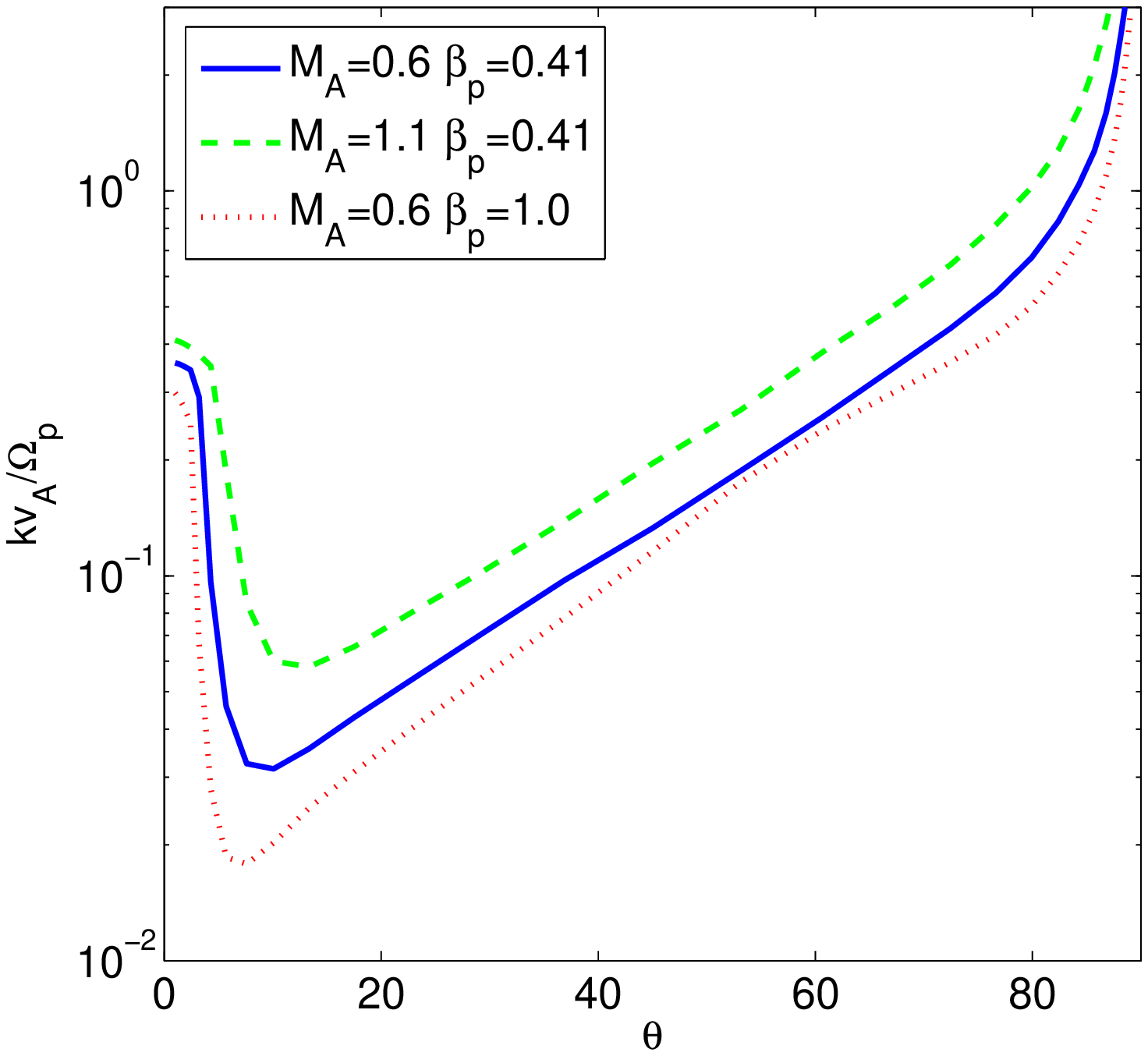}
\end{center}
\caption{{\it Left}: Green and red lines show the sample contour of cascading rate and damping rate.
When the two contours with equal value cross (the two contours cross at $45^{\rm o}$ in this example), 
we obtain the critical $k_c$, which roughly labels turbulence spectral cut-off in different directions. 
The cascading rate increase at large angles and vice versa, 
so the $k_c$ goes above the damping rate contour at large angles and below the
contour at small angles. This variation of $k_c$ over angle 
produces an extended cut-off region in $k$ space when the
turbulence spectrum is integrated over angle (see Figure \ref{W-k-f_spec.ps}). 
{\it Left}: The resulting $k_c$ curve for different conditions. Solid, dash and dotted lines show
the case where ($M_A=0.6$, $\beta_p=0.41$), ($M_A=1.1$, $\beta_p=0.41$) and 
($M_A=0.6$, $\beta_p=1.0$) respectively. Apparently, the $k_c$ contour is shifted to high $k$s at high turbulence 
intensity and lower $k$s at high damping rate.
}
\label{kc_contour.ps}
\end{figure}

The numerical result of steady state turbulence spectrum for isotropic diffusion tensor is highly anisotropic 
(see Figure \ref{taus_cut.ps}). 
To understand the result from cascading and damping, we perform simulation without thermal damping 
(instead of including thermal damping at proper $k$, we set an artificial isotropic cut-off at very large $k$). 
In such a case, the injected turbulence 
evolves into a quasi-isotropic Kolmogoroff spectrum in MHD region and breaks slightly beyond that. 
The wave cascading rate, as we discuss on MHD turbulence in \S\ \ref{general},
is large near perpendicular direction and small when otherwise (see Figure \ref{taus_cut.ps} and Figure \ref{kc_contour.ps}). 
Meanwhile, the wave damping is weaker near perpendicular direction than in the parallel direction.
This anisotropic damping, when combined with anisotropic cascading, makes the Alfv\'{e}n wave spectra cut off
at different magnitude of $k$ on different angles, and the cut-off region extends
over one order of magnitude in wave vector space when integrated over over the $k$-sphere. 
As shown in Figure \ref{W-k-f_spec.ps}, the extended cut-off region has a spectrum close to 
a power law with index $3\sim4$. This spectrum suggests us to use it in explanation of the observed broken 
power law spectrum in solar wind Alfv\'{e}nic turbulence. To compare with the observation 
we integrate the steady state turbulence over wave vector space to obtain the spectrum in Doppler-shifted frequency space
of the spacecraft (Leamon et al. 1999):
\begin{equation}
\label{transf}
F(f) = \int {\cal W}({\bf k})
	\delta \left(\frac{1}{2\pi}\left({\bf k} \cdot {\bf V}_{SW} + \omega({\bf k}) \right) - f 
			\right)
	d{\bf k} 
\end{equation}
where $f$ is the spacecraft-frame frequency, $F(f)$ is observed spectrum and $\delta(...)$ is the Dirac delta function.  
As shown in Figure \ref{break.ps}, after this Doppler-shifting integral, the extended cut-off region becomes a 
power-law-alike spectrum. And this spectrum provides a good fit to the spectrum of the Alfv\'{e}nic oscillation in 
solar wind observed by Leamon et al. 1998. 
Furthermore, since this broken power law is a direct result of the extended cut-off region in wave vector space, we can approximate
the break frequency $\nu_{bf}$ with the starting point of this cut-off region. If we define the anisotropic cut-off points in wave vector
space as ${\bf k_{c}}(\theta)$, then the observed break frequency can be roughly estimated by
\begin{equation}
\label{nu_bf}
\nu_{bf}\sim \min_{\theta}({\bf k_{c}}(\theta)\cdot{\bf V_{SW}})
		=\min_{\theta}(k_{c,||}V_{SW}\cos(\Theta_{BV})+k_{c,\perp}V_{SW}\sin(\Theta_{BV}))
\end{equation}
We can see from Figure \ref{kc_contour.ps} that $k_{c}$ is much large at near perpendicular
direction, thus we can expect that
${\bf k_{c}}(\theta)\cdot{\bf V_{SW}}$ increase dramatically with the increase of $\Theta_{BV}$.
However, the $\min_{\theta}(...)$ in the formula mixed up the simple relationship and 
we obtain a weakly dependence of the $\nu_{bf}$ on $\Theta_{BV}$ (see Figure \ref{ang_dep.ps}). 
This result qualitatively agrees with two similar cases observed by Leamon et al. (1998),
where all parameters are 
close except for the angle of solar wind. It may come to attention that in the $\Theta_{BV}=87^{\rm o}$
case, the observation, the second index of the broken power law appears much harder than the 
$\Theta_{BV}=23^{\rm o}$ case, which is on the contrary to our numerical result. We argue that
in the $\Theta_{BV}=87^{\rm o}$ case, the second power law segment is too close to 
the background noise to fit an accurate index number. Thereafter, we only need to take the 
break frequency into serious consideration.
We further discuss this effects of noise saturation on fitted index number 
in \S\ \ref{observables}

\begin{figure}[htb]
\begin{center}
\centerline{\plotone{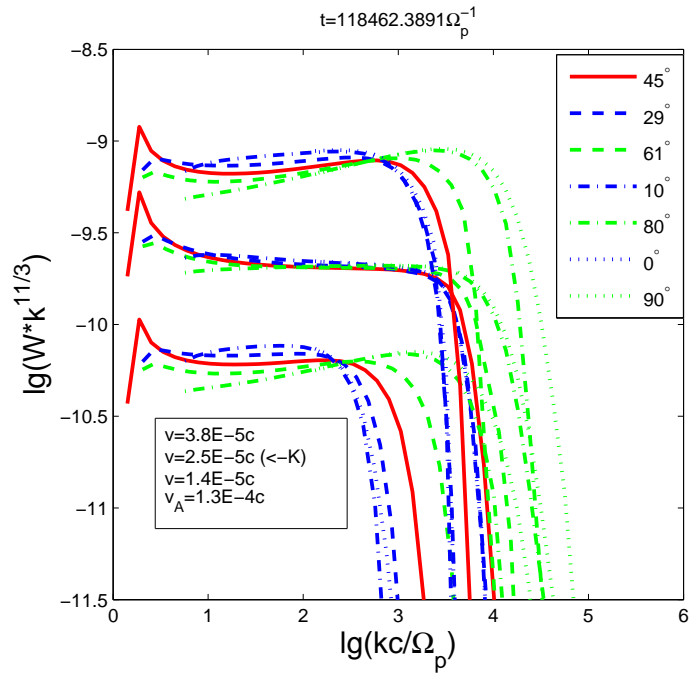}}
\end{center}
\caption{Spectral cut-off in different directions. A Kolmogoroff (Homogeneous) cascading with anisotropic
damping is simulated (the middle bunch of lines) to compare with result our diffusion tensor model.}
\label{testani.ps} 
\end{figure}

\begin{figure}[htb]
\begin{center}
 \includegraphics[width=0.3\textwidth]{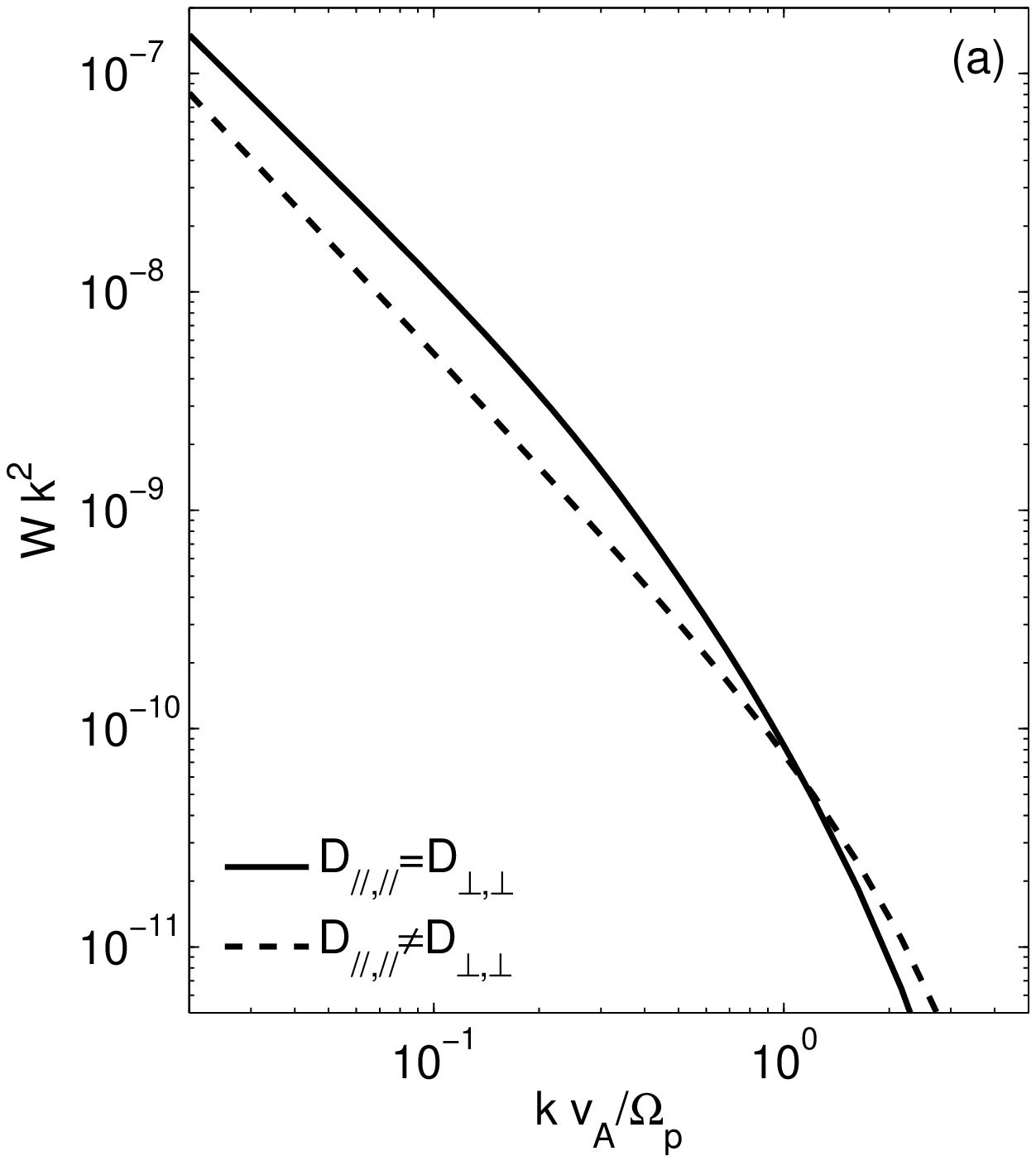}
 \includegraphics[width=0.3\textwidth]{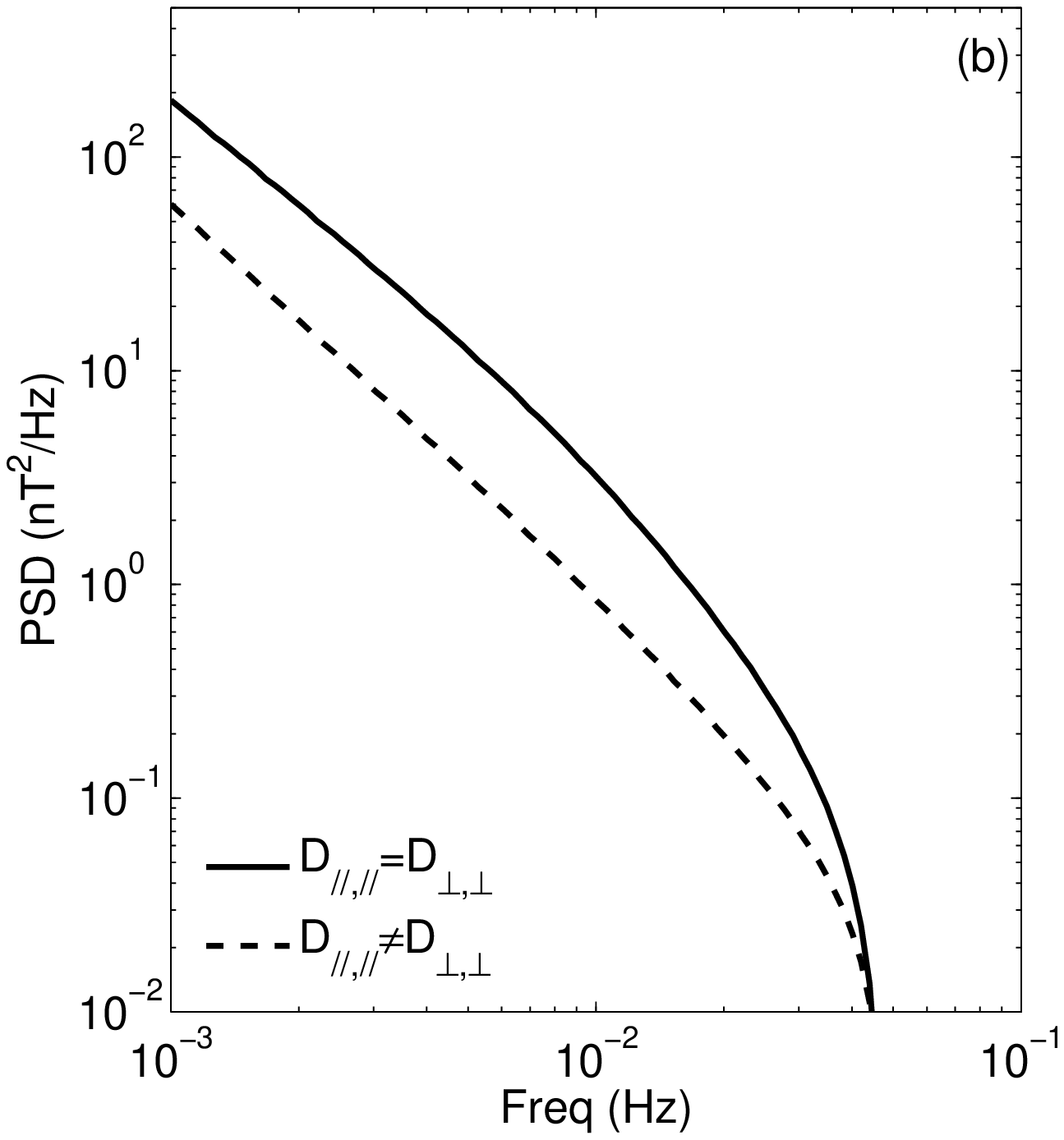}
 \includegraphics[width=0.3\textwidth]{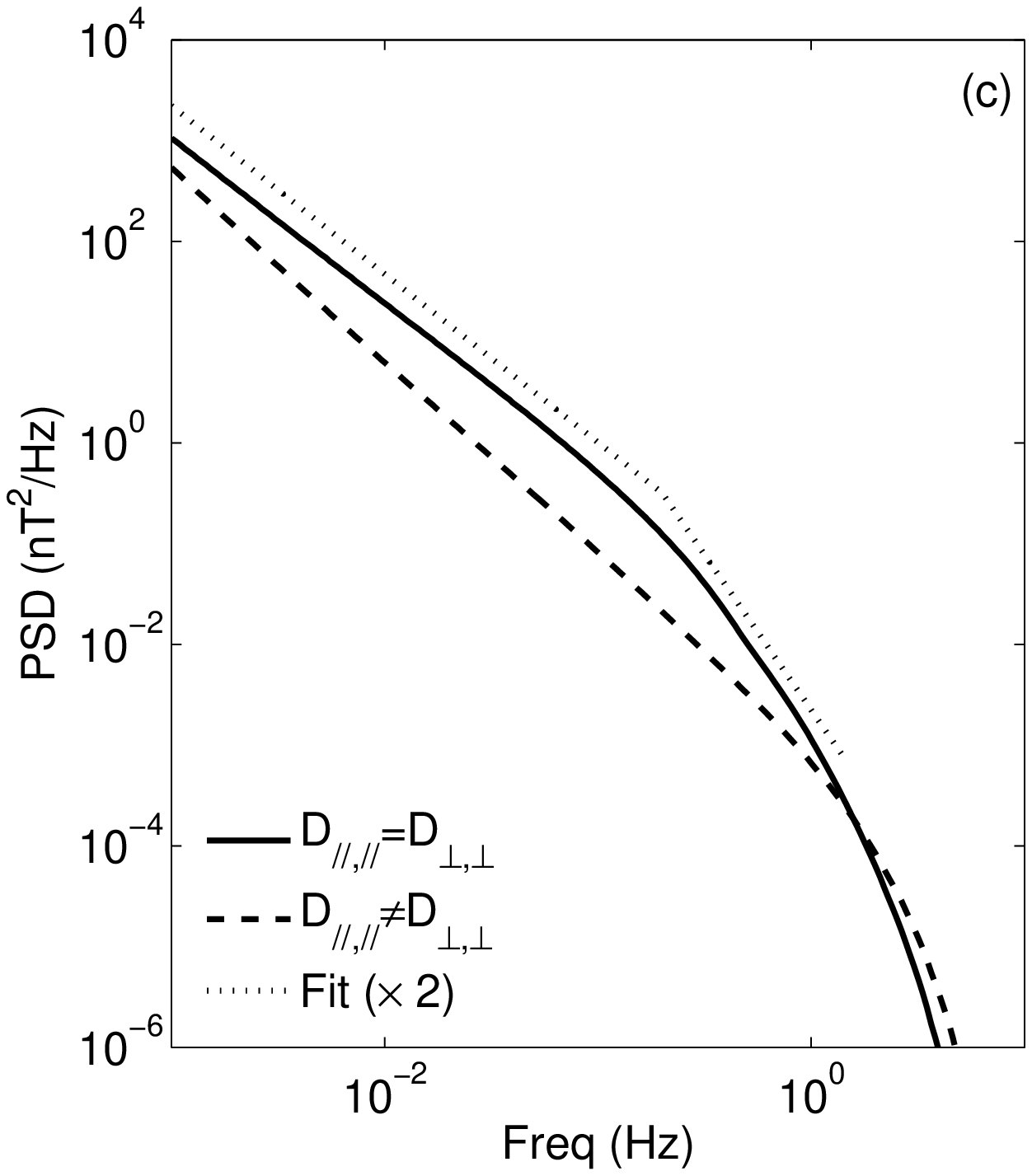}
\end{center}
\caption{
{\it Panel (a)}: Wave turbulence spectrum as function of k after integration over angle. The result from 
isotropic diffusion tensor model (red) steepens slowly and the cut-off region extends about one order 
of magnitude in wave vector space; the result from anisotropic tensor model (blue)
cuts off rapidly around $~1/\rho_p$ ($vA\approx v_{th,p}$ in this simulation). 
The latter spectrum agrees with Howes et al. 2007.
{\it Panel (b)}: Wave turbulence spectrum as function of wave frequency when solar wind Doppler-shift 
is absent. Both spectra show a Kolomogoroff index and cut off at Hellium gyrofrequency 
(i.e. the upper limit of cold plasma Alfv\'{e}n dispersion surface). 
{\it Panel (c)}: The spectra in Doppler shifted space-craft frequency for two $Dij$ models. The 
result of isotropic tensor model can be fit into a broken power law spectrum, with 
$\gamma_1=-1.67$, $\gamma_2=-3.1$ and break frequency $\nu_{bf}=0.2$Hz.
}
\label{W-k-f_spec.ps}
\end{figure}

\begin{figure}[htb]
\begin{center}
\centerline{\plotone{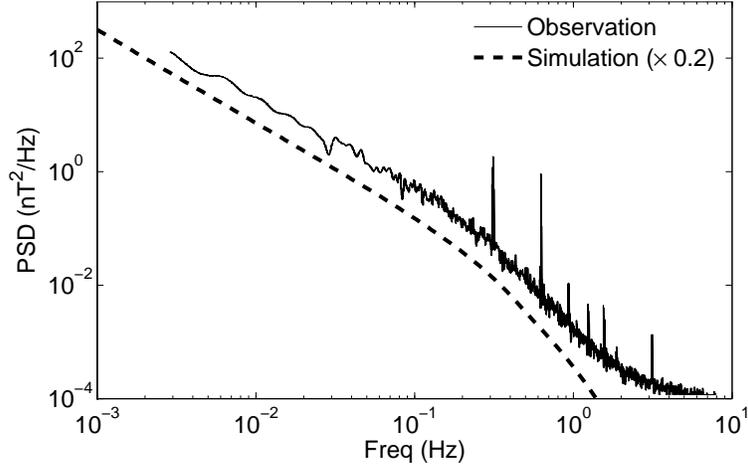}}
\end{center}
\caption{
Observed interplanetary turbulence power spectrum (solid) vs. simulation results (dash). 
The data are extracted from Figure 1. by Leamon et. al. (1999). For this 
observation period, $B=6.3$nT, $\beta_p=0.48$, $\Theta_{BV}=38^{\rm o}$, $V_{SW}=517$km/s, 
$v_A$ is unknown (we use 40km/s for our simulations).
With these parameters as input, our isotropic model ($D_{||,||}=D_{\perp,\perp}$) provides 
an excellent fit to the observation. Leamon et. al. (1999) fit the observation with a broken power law
with indices $\gamma_1=-1.67$, $\gamma_2=-2.91$ and break frequency at $\nu_{bf}=0.235$Hz. 
Our simulated spectrum, when fit to a broken power law down to $10^{-3}$PSD, provides 
$\gamma_1=-1.67$, $\gamma_2=-2.97$ and break frequency $\nu_{bf}=0.200$Hz.)
}
\label{break.ps} 
\end{figure}

\begin{figure}[htb]
\begin{center}
 \includegraphics[width=0.45\textwidth]{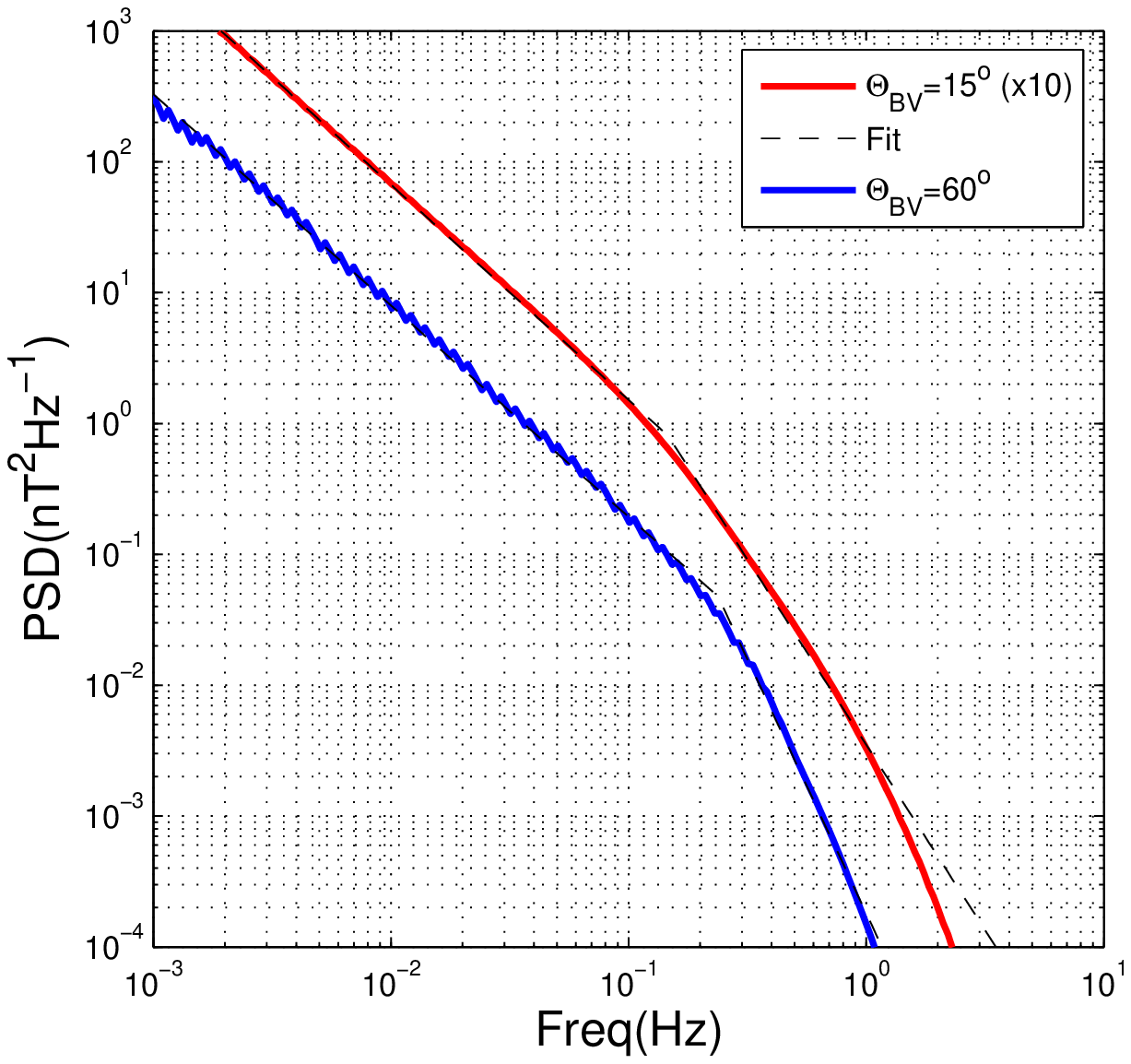}
 \includegraphics[width=0.45\textwidth]{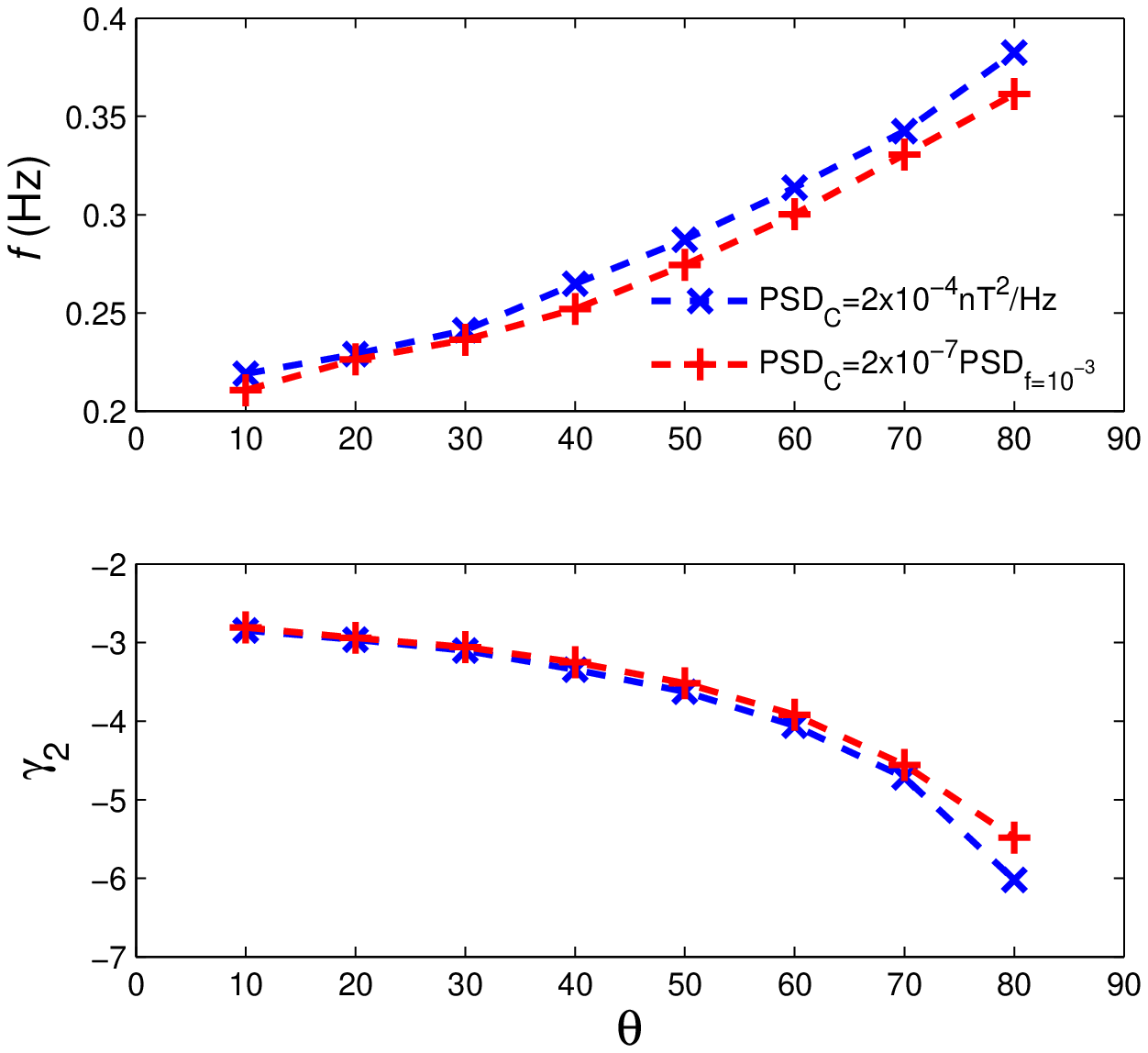}
\end{center}
\caption{
Numerical simulation result for angle dependence of break frequency. {\it Left}: The numerical simulation of 
spacecraft frequency spectra at different solar wind to mean magnetic field angle, $\Theta_{BV}$, 
with all other parameters equal ($M_A=0.6$ and $\beta_p=0.41$, see Figure \ref{break.ps} for values of specific
observables). Broken power law fits provides break frequency, $\nu_{bf}=0.15$
for $\Theta_{BV}=15^{\rm o}$ and $\nu_{bf}=0.24$ for $\Theta_{BV}=60^{\rm o}$.
{\it Right}: Setting observational turbulence power cut-off as ${\rm PSD_C} = 5\times10^{-4}({\rm nT^2Hz^{-1}})$ (blue) 
or ${\rm PSD_C} = 1\times10^{-4}({\rm nT^2Hz^{-1}})$ (red), we simulate the relation between 
break frequency and solar wind angle in the upper panel and relation between after-break-index and
solar wind angle in the lower panel. (Note that when observational noise and saturation is taken into 
account, the index after break may not be accurately measured.) Numerical simulation shows that the 
break frequency weakly depends on the solar wind angle. The other parameters are fixed at the same value
as in {\it Left} panel.  
}
\label{ang_dep.ps} 
\end{figure}

We emphasize that this broken power law spectrum requires anisotropy of cascading as well as damping. 
Figure \ref{ang_dep.ps} gives a direct comparison between the ${\bf k_{c}}(\theta)$ and equal $\omega_i$ contour in
wave vector space. Note that the $\omega_i$ contour would mark the cut-off in different direction 
if the cascading was isotropic. Since $\omega_i$ is more isotropic, the cut-off region would not be 
extended enough to make a broken power law spectrum without anisotropic cascading. 
The numerical test with an artificial 
isotropic cascading (we use Kolmogoroff cascading with lower injection rate) confirms our conclusion
(see Figure \ref{testani.ps}). \footnote{By assuming a hard spectral index 
($\gamma_2=5$ in the paper) 
beyond certain $\omega_i$ contour in wave vector space, Leamon et al. (1999) obtained a 
broken power law spectra in frequency space. They argue that thermal damping alone 
can explain the observation. However, our simulation shows 
that the turbulence spectra cut off very steeply when damping rate is large enough. 
Thereafter their assumption is invalid, so does the conclusion.}

\subsection{More Anisotropy by Anisotropic Diffusion tensor}
\label{aniresult}

The non-isotropic diffusion tensor provides a Kolmogoroff index 
spectrum in perpendicular direction only. In other directions, the energy spectrum is a much steeper.
Although the spectral cut-off spreads to a even larger range of $k$ in such a case, the energy in 
parallel direction contributes very little to the $k$-sphere integrated spectrum.
As shown in Figure \ref{W-k-f_spec.ps}, there is no extended cut-off region for this model. As a result,
when integrated into Doppler-shifted frequency space, the spectrum after break is much more steeper 
than observation range (with index $~7$) even we fit it as a power law. 
This disagreement with observation
is also found in many numerical or analytical studies (Li 19XX, Howes et al. 2007).
\footnote{Howes et al (2007) suggest that the spectral cut-off, when combined with a observational 
noise saturation, may appear to be a power law spectrum. In this paper we stick on the 
broken power law conclusion observed by Leamon et al. (1998) and make a direct comparison between observation
and our theory.} 
In a word, 
a simple thermal damping can not produce the observed broken power law in a 1-D cascading model 
(or 2-D if we take the two orthogonal direction in perpendicular direction separately). 
When most of the turbulence energy is on perpendicular or close to perpendicular direction,
we can only refer to change of cascading rate itself to produce a broken power law. 
The KAW, with a dispersion relation diverges from the cold plasma or MHD approximation, provides
a different cascading rate. However, Howes et al. (2007) show that the spectral index of KAW is
only -7/3, much less than second index of the observed broken power law. Furthermore, 
KAW only starts to appear at $k_{\perp} \rho_p \approx 1$ ($\rho_p$ is proton Larmor radius), 
which is Doppler-shifted to a spacecraft frequency 
\begin{equation}
\nu \approx {\bf k_{perp}}\cdot V_{SW} \approx (k_{\perp}\rho_p {V_{SW,\perp}\over v_A} \beta_p^{-1/2})\nu_p
\end{equation}
where $\nu_p$ is proton gyrofrequency. In typical Solar Wind conditions, where $\beta_p<1$ or $\sim 1$ and
$V_{SW}\sim 10 v_A$, this frequency is 10 times the proton gyrofrequency or over. Whereas most of the 
observations (Leamon et al. 1998 \& 1999, Bale et al. 2005) contains a break frequency smaller than 
$10\nu_p$. Although the available observation results are not complete enough to rule out KAW as the 
reason for the broken power law, a different mechanism other than KAW is suggested.

To generalize, we are unable to reproduce the observed broken power law spectrum with our anisotropic 
diffusion tensor model. We also show that this difficulty persists on any model that producing a highly 
anisotropic turbulence spectrum. The isotropic diffusion tensor (with non-isotropic cascading rate), 
on the contrary, provides a good fit to the observation. A further observational evidence is the weak
angle dependence of the break frequency (Leamon et al. 1998), which can be very difficult to explain for 
a highly anisotropic spectrum but can easily fit into the isotropic diffusion tensor model. Thereafter,
in the following discussion and observational prediction, we use the isotropic diffusion tensor model only.

\subsection{Break frequency and observables}
\label{observables}

In our model, the broken power law is produced by the wave spectral cut-off at different magnitude of $k$ 
in different direction, hence the breakpoint in spacecraft-frame frequency space corresponds to the 
cut-off in wave vector space indicated by Formula (\ref{nu_bf}).
This model allows us to draw some conclusions on the relation between spectral breakpoint and the 
observable parameters.

\begin{figure}[htb]
\begin{center}
 \includegraphics[width=0.45\textwidth]{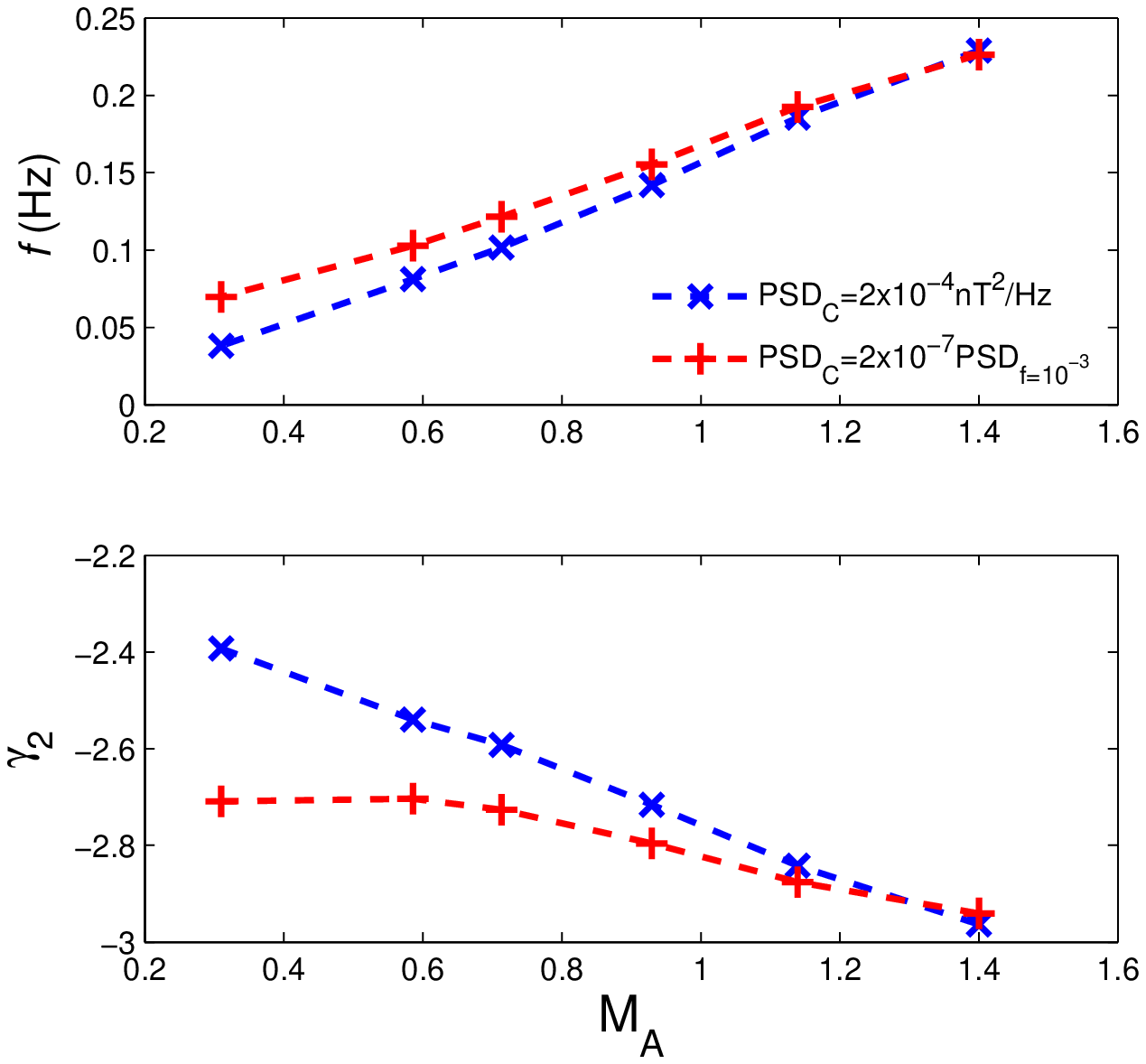}
 \includegraphics[width=0.45\textwidth]{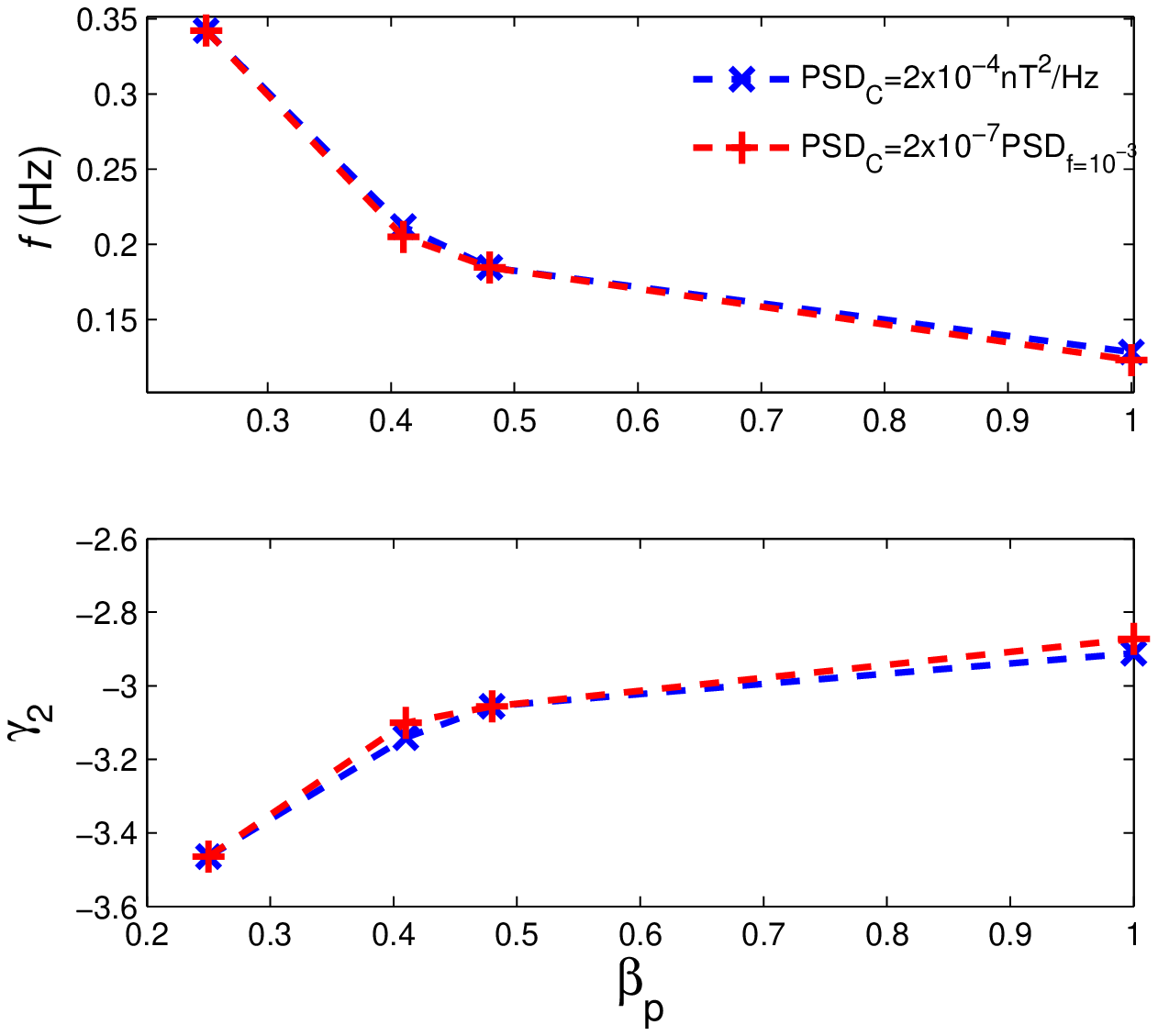}
\end{center}
\caption{
{\it Left}: Same as Figure \ref{ang_dep.ps}, but dependence on the Alfv\'{e}n Mach number. Other parameters
are fixed at the same value as in Figure \ref{break.ps}, i.e. $\Theta_{BV}=38^{\rm o}$, $\beta_p=0.41$.
{\it Right}: Dependence on plasma beta. Other parameters
are fixed at the same value as in Figure \ref{break.ps}, i.e. $\Theta_{BV}=38^{\rm o}$, $M_A=0.6$.
}
\label{MA_dep.ps} 
\end{figure}

We categorize the observables into four independent parameters, Alfv\'{e}n Mach number $M_A$, 
plasma beta $\beta_p$, $V_{SW}$, and $\Theta_{BV}$. Obviously, a faster solar wind will give a 
higher break frequency. This relation is defined by the Doppler-shift term
${\bf k}\cdot{\bf V_{SW}}$ in Equation (\ref{transf}) rather than depends on turbulence model.
Instead, we are interested in relations that are turbulence model dependent as below.
First, since the cut-off (or break in observation) is produced by turbulence damping overwhelming 
its cascading, the relative value between 
turbulence intensity and thermal damping rate will affect the breakpoint the most. We can expect 
that a high
Alfv\'{e}n Mach number will give a high break frequency and high plasma beta will reduce break frequency.
Our numerical simulation in Figure \ref{MA_dep.ps} confirms these relations. 
Note that since the cascading rate
still contains one undetermined free parameter, the cascading constant $C$ in Equation (\ref{DijMHD}), we can only
predict the $\nu_{bf}$-$M_A$ or $\nu_{bf}$-$\beta_p$ relation rather than provide the exact value of break frequency
from observed turbulence parameters. 
Although it is interesting to notice that $C=1$ gives a very close match to
the observation made by Leamon et. al. (1999) in both break frequency and turbulence intensity, we  
\footnote{In principle, we are able to determine the
cascading constant by fitting our theoretical breakpoint with this observation. However, we are not able to 
obtain all the observational parameters for the individual case from Leamon et. al. (1999). 
We pick a typical Alfv\'{e}n wave
velocity in solar wind as $v_A=40 {\rm km/s}$, and use this value for all our simulations.}
Secondly, the projection effect from $\Theta_{BV}$ also affects the break frequency. It is obvious that
the break frequency increase at larger projection angles.
However, the angle dependence is not as critical as predicted by any 2-D (wave energy cascades only to perpendicular direction) 
turbulence model. With 3-D wave spectrum and cut-off wave vector $k_c$ in our model with isotropic diffusion tensor, 
the minimum value of 
${\bf k_c}\cdot{\bf V_{SW}}$ depends on both the shape of $k_c$ contour and the angle of solar wind. 
A more isotropic cut-off wave vector will give less dependency on the angle of solar wind. On the other hand,
if most of the wave energy resides in perpendicular direction, as suggested by critical balanced spectra, 
the $k_c$ contour will be highly prolonged in perpendicular direction and the angle of solar wind 
will strongly affect the break frequency. The observation by Leamon et al (1998, Figure 6) fits better
with a more isotropic model as ours but further observation is required to rule out any one. 

The spectral index after the break, on the other hand, is also related to these observable parameters as 
we show in Figure \ref{ang_dep.ps} and \ref{MA_dep.ps}. However, since the broken power law is 
only an observational effects of the extended cut-off region rather than a mathematical secondary power law spectrum, 
the fitted index number highly depends on the low energy cut-off of the observation, and the noise saturation
also effects the result (Howes, et. al. 2007). Thereafter, we do not attempt to compare theory with 
observations unless further detector can measure a definite result on the index number.

\section{Conclusion}
\label{dis}

With diffusion approximation, we are able to construct a diffusion tensor discribing Alfven-cycloton 
turbulence cascade. This cascade model, when coupled with thermal damping, provides a broken
power law spectrum after integration, which matches the observed solar wind plasma turbulence 
spectrum by Leamon et. al. (1999). By comparing the model with observation, we conclude that
the observed broken power law spectrum comes from a quasi-isotropic turbulence spectrum that 
cuts off at different $k$ in different direction. Based on this model, we predict the observed 
break frequency is propotional to Alfven Mach number, and antipropotional to plasma beta. 
This result can be subjected to further observational tests.

\acknowledgments

This work was carried out under the auspices of the National Nuclear Security Administration of the U.S. 
Department of Energy at Los Alamos National Laboratory under Contract No. DE-AC52-06NA25396. This 
research was partially supported by NSF grant ATM-0312344, NASA grants NAG5-12111, NAG5 11918-1 (at 
Stanford).

\appendix
\begin{center}
    {\bf APPENDICES}
\end{center}
\section{The Exact Solution of Cold Plasma Dispersion Surface}
\label{cold}
The equation for cold Plasma dispersion surface can be constructed through solving Maxwell's
equation for plane waves (Stix 1962). A nontrivial wave solution to the equation requires the 
refractive index $\bf{n}={\bf{k}c \over \omega}$ to satisfy equation,
one obtain the wave equation,
\begin{equation}
{\bf n}\times({\bf n}\times{\bf E})+ {\sf K} \cdot {\bf E}=0
\label{waveeq.eq}
\end{equation}
where $\sf{K}$ is the dielectric tensor defined by 
\begin{equation}
{\bf J}-i\omega\epsilon_0{\bf E} \equiv -i \omega \epsilon \sf{K}\cdot {\bf E}. 
\end{equation}
By defining $P, R, L, S=1/2(R+L), D=1/2(R-L)$ as a function of $\omega$, and $A, B, C$ terms 
as a function of $P, R, L$ and $\theta$, Stix (1962) was able to write all the wave solution in
the form $\displaystyle n^2={B \pm F \over 2A}$. Although this solution is segmented and 
has may poles at particles' cyclotron frequency, each physical mode of dispersion surface is both 
continuous and smooth almost everywhere. In the following, we express these continuous modes by 
combining with segmented solutions of the equation (\ref{n2eq}), i.e.

Alfv\'{e}n branch,
\begin{equation}
k=\pm \omega \sqrt{B-F \over 2A} \qquad \omega\in[0,\Omega_{\alpha})
\end{equation}
$R$ and $L$ term reach their first pole ($\Omega_{\alpha}$), and we have $\lim_{\omega\to\Omega_{\alpha}^-}k=\infty$.
This point is the end of Alfv\'{e}n branch. 

Fast branch,
\begin{displaymath}
k=\left\{ \begin{array}{ll} \displaystyle\pm \omega \sqrt{B+F \over 2A} & \omega\in[0,\Omega_{\alpha}^-]\\
\displaystyle\pm \omega \sqrt{B-F \over 2A} & \omega\in[\Omega_{\alpha}^+,\Omega_{p})\\ \end{array} \right.
\end{displaymath}
Similarly, Fast branch ends at $\Omega_p$ with $k\to\infty$.
Although $\Omega_{\alpha}$ is a pole for $R$ and $L$, $\lim_{\omega\to\Omega_{\alpha}^-}{B \pm F \over 2A}=
\lim_{\omega\to\Omega_{\alpha}^+}{B \pm F \over 2A}\neq\infty$.
To understand the switching sign at $\Omega_{\alpha}$, I study the simplified problem at $\theta=0$. In 
such a case $F=2\sqrt{P^2D^2}=2\mid PD \mid$ and I get
\begin{equation}
k=\pm \omega\left(S \pm {\mid PD\mid \over P}\right)^{1/2}
=\pm \omega\left(S \pm sign(PD)D\right)^{1/2}
\end{equation}
Since $P<0$ for all $\omega\in[0,1]$, and $D$ switch sigh at $\Omega_{\alpha}$, the discontinuity is only
introduced by our attempt to write $k(\omega)$ in an explicit form. By switching sign in ${B \pm F \over 2A}$ 
at $D=0$, I can follow the continuous dispersion surface. 

Whistler branch
\begin{displaymath}
k=\left\{ \begin{array}{ll} \displaystyle\pm \omega \sqrt{B+F \over 2A} & \omega\in[\omega_1,\Omega_{p})\\
\displaystyle\pm \omega \sqrt{B-F \over 2A} & \omega\in(\Omega_{p},\omega_{e})\\ \end{array} \right.
\end{displaymath}
Whistler branck starts at $\omega_1$, which solves the equation $R(\omega)=0$. 
$\omega_1\approx0.585\Omega_p$ (very unsensitive to density and field strength) by numerically
solving the equation. $P=0$ at $\omega_e$, the fomular reach the pole in parallel direction 
($\theta=0$), and the Whistler branch ends at the electron Langmuir oscillation.

\end{document}